\documentclass[aip,cha,reprint,amsmath,amssymb,floatfix]{revtex4-1}
\usepackage{color}
\newif\ifshowrevision
\showrevisionfalse
\DeclareRobustCommand{\revadd}[1]{\ifshowrevision{\color{black}#1}\else#1\fi}
\newif\ifshownewrevision
\shownewrevisiontrue
\DeclareRobustCommand{\revnew}[1]{\ifshownewrevision{\color{black}#1}\else#1\fi}

\usepackage{iftex}
\usepackage{pifont}
\newcommand{\cmark}{\ding{51}}
\newcommand{\xmark}{\ding{55}}
\ifPDFTeX
\usepackage[T1]{fontenc}
\usepackage[utf8]{inputenc}
\usepackage{textcomp}
\usepackage{mathptmx}
\else
\usepackage{fontspec}
\fi
\usepackage{array}[=2016-10-06]
\usepackage{graphicx,booktabs,tabularx}
\usepackage[unicode,hidelinks]{hyperref}
\hypersetup{pdftitle={Learning effective stochastic predator and prey dynamics across physical and biological systems from limited data}}
\usepackage{etoolbox}
\begin{document}
\pagestyle{plain}
\title{Learning effective stochastic predator and prey dynamics across physical and biological systems from limited data}
\author{J. C. Huang}
\affiliation{School of Physical and Mathematical Sciences, Nanyang Technological University, Singapore 637371, Singapore}
\author{Z. S. Qu}
\email{zhisong.qu@ntu.edu.sg}
\affiliation{School of Physical and Mathematical Sciences, Nanyang Technological University, Singapore 637371, Singapore}
\author{R. Varennes}
\affiliation{CEA, IRFM, F-13108 Saint-Paul-Lez-Durance, France}
\author{X. Garbet}
\affiliation{School of Physical and Mathematical Sciences, Nanyang Technological University, Singapore 637371, Singapore}
\affiliation{CEA, IRFM, F-13108 Saint-Paul-Lez-Durance, France}
\author{C. G. Wang}
\affiliation{School of Physical and Mathematical Sciences, Nanyang Technological University, Singapore 637371, Singapore}
\author{C. Guet}
\affiliation{School of Physical and Mathematical Sciences, Nanyang Technological University, Singapore 637371, Singapore}
\author{D. Niyato}
\affiliation{College of Computing and Data Science, Nanyang Technological University, Singapore 639798, Singapore}
\author{K. Lim}
\affiliation{School of Physical and Mathematical Sciences, Nanyang Technological University, Singapore 637371, Singapore}
\author{V. Grandgirard}
\affiliation{CEA, IRFM, F-13108 Saint-Paul-Lez-Durance, France}

\begin{abstract}
\revnew{Constructing an effective low-dimensional stochastic model from sparse observations requires separating deterministic drift from state-dependent fluctuations when the governing interaction law is unknown. We formulate an adaptive neural stochastic differential equation with topology constraints (TC-AdaNSDE) for two-variable predator--prey dynamics. To regularize the inverse problem when data are scarce, the model combines flexible drift and correlated diffusion fields with density-adaptive local Kramers--Moyal moment estimates and soft phase-space constraints reflecting boundary saddle behavior near extinction and local stability near coexistence. Using a known nonlinear stochastic system, we first show that short-time trajectory error cannot assess diffusion accuracy on its own and must be paired with distributional measures of one-step transitions and long-time phase-space occupancy. We then apply this framework to macroscopic turbulence and zonal flow energies reduced from Hasegawa--Wakatani plasma simulations and to short ecological time series. In the sparse-data regime, TC-AdaNSDE reproduces both short-horizon evolution and long-time invariant energy distributions where deterministic reductions collapse and unconstrained stochastic fits diverge, while recovering physically interpretable drift and fluctuation fields across both physical and biological systems.}
\end{abstract}
\maketitle
\thispagestyle{plain}

% Chaos requires a lead paragraph for nonspecialists. Draft for author review.
\begin{quotation}
\revnew{Predator--prey feedback governs cyclic dynamics across disciplines, from interacting biological populations to the exchange of energy between microturbulence and shear flows in magnetized plasmas. When these high-dimensional systems are observed through only two macroscopic variables over short records, unresolved degrees of freedom act as effective random forcing, and standard parametric equations are often too restrictive. Conversely, purely data-driven models require extensive observations to separate systematic drift from fluctuations and can easily become unstable where phase space is poorly sampled. Here we show that combining local increment statistics with qualitative phase-space structure, specifically, growth and decay tendencies near extinction and a restoring tendency near coexistence, allows a flexible stochastic model to be inferred reliably from very short records. Tests on reduced plasma turbulence energies and empirical population time series demonstrate that these qualitative dynamical priors yield stable models that capture both short-term evolution and statistical distributions from only a few observed cycles.}
\end{quotation}

\section{Introduction}\label{sec:introduction}

Predator and prey feedback provides a common description of nonlinear interactions in biological and physical systems. In ecology, prey sustain predator growth while predators deplete the prey. In magnetized plasmas, turbulence drives zonal flows, whose shear in turn suppresses the turbulence \cite{Diamond2005}. This shared feedback suggests a description with few variables, but does not specify a common explicit interaction law. Population dynamics depend on environmental conditions and unresolved biological structure, while turbulence and flow energy exchange depends on spatial fields that two energy variables cannot resolve. The resulting observations can retain recognizable predator and prey behavior while exhibiting irregular fluctuations. Learning a useful stochastic description therefore requires both flexibility in the interaction law and guidance about the dynamics that should be retained.

Our earlier work \cite{huang2026} showed that a stochastic predator and prey model could be learned from the turbulence and zonal flow energies recorded in Hasegawa--Wakatani (HW) simulations. The aim was to describe the irregular exchange of energy using just these two quantities, without tracking the full spatial plasma dynamics. Neural networks learned their average evolution and fluctuations, while known features of the plasma interaction helped guide the model where data were limited. This demonstrated the value of combining observations with physical knowledge in a particular system. The present study asks how far that approach can extend when such detailed physical knowledge is unavailable: can qualitative features shared by predator and prey interactions provide enough guidance to learn useful stochastic models in other physical and biological settings?

Limited observations make this question particularly important. Short population records may contain only a few oscillations, with uneven coverage of the two variable state space and few transitions near any given state. Reconstructing stochastic dynamics from data relies on estimating average changes and fluctuations near each state, as reviewed by Friedrich et al. \cite{friedrich2011}, while sparse coverage makes these estimates uncertain. A prescribed parametric equation can reduce the amount of data needed for fitting, but may exclude relevant nonlinear interactions. Neural stochastic differential equations (SDEs) use neural networks to represent both the average motion and random fluctuations \cite{tzen2019,li2020}, yet many different models may fit the same short record. Individual observed changes cannot reliably distinguish systematic motion from stochastic variability, and poorly sampled states provide little information about how the system behaves there. The challenge is to use the statistical information and qualitative dynamical knowledge available without fixing an interaction formula in advance.

Here we develop an adaptive neural SDE with topology constraints (TC-AdaNSDE) that extends the earlier approach through more general dynamical guidance and a fitting procedure designed for sparse, uneven observations. The model learns average changes and fluctuations together, pooling transitions from nearby states to improve local statistical estimates. The neighborhoods used for this pooling expand where observations are sparse and contract where they are dense. The model also learns how fluctuations in the two variables are correlated, allowing their coupling to differ between systems and across states. Its structural guidance uses population or energy magnitudes to scale changes and favors two qualitative behaviors: prey growth and predator decline near extinction, and a tendency to return toward coexistence after a small disturbance. These preferences guide the learned dynamics without prescribing a particular interaction equation. The advance beyond the plasma study is therefore a framework that can use limited observations together with this partial knowledge of predator and prey behavior, with its usefulness tested across systems with different interaction mechanisms.
%These are modeling assumptions for the class of interactions considered, rather than universal properties of predator--prey systems. 
%Here ``topology-constrained'' refers to this local guidance of the drift; it does not imply a guarantee of global phase-space topology. 
%Figure~\ref{fig:learning_workflow} summarizes how the statistical and structural information enter the model.

\revadd{We evaluate the framework on a simulated nonlinear system with a prescribed equation, plasma simulations, and short biological records. The controlled example varies the number of fitting transitions at fixed observation intervals to examine when partial dynamical guidance helps. It distinguishes accurate local prediction and transition statistics from the state distributions generated by autonomous simulations.} Energies extracted from HW simulations test whether two observed quantities can usefully describe an interaction that depends on complex spatial dynamics. Rotifer and algae records \cite{blasius2020} and Least Killifish and Eastern Mosquitofish observations \cite{gatto2024} test learning from measured population data whose governing equations are unknown. We assess both the accuracy of predicted changes and how closely simulated fluctuations reproduce the observed distributions. In the plasma tests, the benefits of TC-AdaNSDE are clearest when few transitions are retained for fitting: it combines competitive prediction over short intervals with closer agreement in the distribution of energy states than the principal neural SDE comparisons. Across the short biological records, it gives a consistently strong balance of these measures. Together, these results support the use of partial predator and prey knowledge to learn useful stochastic descriptions across different systems, within the range of states and time scales evaluated.

This paper is organized as follows. Section~\ref{sec:related_work} reviews related approaches to learning stochastic dynamics. Section~\ref{sec:methodology} presents TC-AdaNSDE and explains how observations and qualitative dynamical knowledge guide its training. Section~\ref{sec:assessment_comparison_models} defines the assessment criteria and the comparison models. Section~\ref{sec:known_equation_controls} \revadd{presents the nonlinear control with different training sizes,} Section~\ref{sec:turbulence_zonal_flow_dynamics} examines the model of turbulence and zonal flow energies, and Section~\ref{sec:biological_applications} reports the biological results. Section~\ref{sec:discussion_limitations} discusses the findings and limitations, and Section~\ref{sec:conclusion} concludes the paper.

% \begin{figure*}[!htbp]

% \centering

% \includegraphics[width=\linewidth,height=0.68\textheight,keepaspectratio]{figures/fig1_method_schematic.png}

% \caption{\label{fig:learning_workflow}How limited observations inform an effective stochastic model. The workflow connects the paper's two requirements: flexibility for an unknown interaction law and guidance for coefficients that sparse observations weakly constrain. Nearby transitions supply local statistical targets, while positive-state scaling and extinction/coexistence constraints guide the neural fit. The output is a continuous stochastic description evaluated against held-out observations. The traces illustrate the method conceptually and are not experimental data.}

% \end{figure*}

\section{Related work}\label{sec:related_work}

Reconstructing stochastic dynamics from observations predates neural SDEs. Early studies inferred drift and diffusion from conditional increment statistics in turbulence and stochastic time series \cite{friedrich1997,siegert1998,friedrich2000}, establishing the approach reviewed by Friedrich et al. \cite{friedrich2011}. Statistical inference for discretely observed diffusions provides a complementary perspective based on likelihoods \cite{craigmile2023}. Gaussian process methods, sparse regression, and stochastic force inference offer different ways to regularize coefficient estimation \cite{batz2018,boninsegna2018,huang2022,frishman2020}. Across these approaches, limited state space coverage and finite observation intervals remain distinct sources of uncertainty. Pooling nearby increments does not itself remove bias from finite observation intervals \cite{ragwitz2001}.

Neural SDEs provide flexible coefficient representations, supported by developments in variational inference, stochastic differentiation, and adversarial learning of trajectory distributions \cite{tzen2019,li2020,kidger2021grad,kidger2021gan}. Kidger \cite{kidger2022} surveys this broader literature. For effective stochastic reduction, Dietrich et al. \cite{dietrich2023} learn neural drift and diffusion from microscopic simulations using losses based on numerical integrators that accommodate snapshot pairs, unequal observation intervals, and partial physical knowledge. Carnazza et al. \cite{carnazza2024} learn stochastic order parameter dynamics using local averages and quadratic variation. These studies establish both neural coarse modeling and statistical targets for coefficient learning. Improved transition approximations, such as DynGMA \cite{zhu2024}, address a complementary numerical aspect of inference.

Recent work also emphasizes stochastic fidelity and sample efficiency. Noise regularization can improve diffusion recovery in variational latent SDEs \cite{heck2025}, while Wasserstein objectives support reconstruction from distributions of jump diffusion processes \cite{xia2024}. These results motivate assessing variability and distributions alongside prediction error, although their formulations differ from the direct transition likelihood used here. Active learning and adaptive sampling based on residuals can reduce data requirements \cite{huang2022,an2024}. TC-AdaNSDE instead adapts the neighborhoods used to pool an existing record, improving local statistical support without acquiring additional observations.

Partial dynamical knowledge offers another source of guidance, through hybrid stochastic models, neural SDEs constrained by physics, and structured Langevin dynamics \cite{haussmann2021,djeumou2023,koop2024}. Building on our stochastic model developed for plasmas \cite{huang2026}, TC-AdaNSDE combines adaptive local moment estimates, correlated neural diffusion, and soft extinction/coexistence constraints. The contribution is this combined construction and its evaluation under limited observations across HW and biological systems. Generalization refers to applying the same strategy with coefficients fitted separately to each system. The constraints guide local drift behavior without prescribing a global interaction law or guaranteeing global topology.

\section{Methodology}\label{sec:methodology}

\begin{figure*}[!htbp]

\centering

\includegraphics[width=\linewidth,height=0.68\textheight,keepaspectratio]{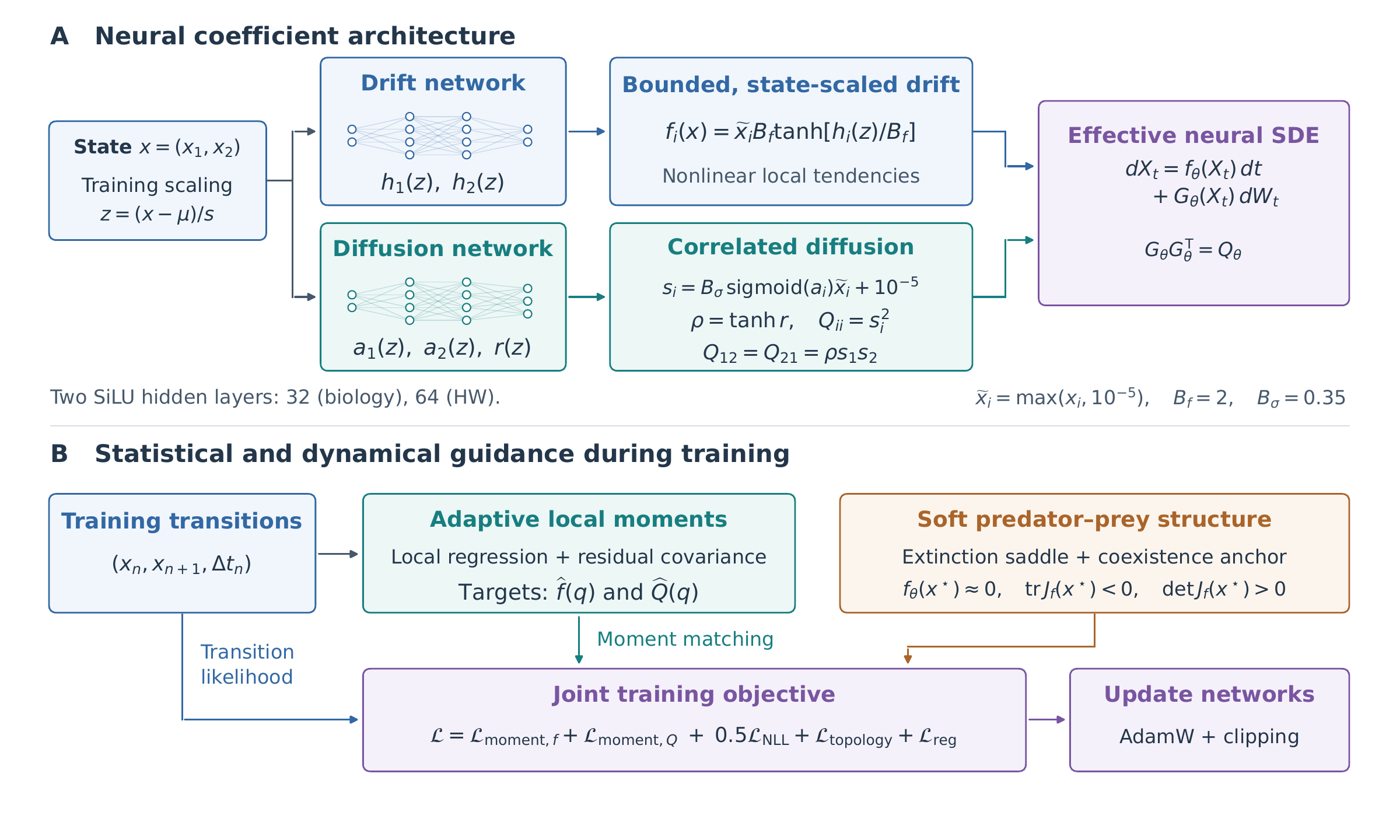}

\caption{\label{fig:model_architecture_training}How the model combines neural flexibility with statistical and dynamical guidance. The upper panel defines the stochastic model. The lower panel shows how limited observations and partial predator and prey structure train it. (A) A standardized two variable state enters separate drift and diffusion networks. Bounded, state-scaled outputs define the drift and a full covariance with state dependent correlation, yielding the effective SDE. Network nodes illustrate connectivity. The widths of the hidden layers are stated below the branches. (B) Observed transitions provide the likelihood and adaptive local moment targets, while soft extinction and coexistence constraints supply dynamical guidance. These terms and coefficient regularization jointly train the networks. Validation selects the neighborhood scale and checkpoint.}

\end{figure*}

We learn a two variable stochastic model by combining three sources of guidance: observed transitions, local averages of nearby transitions, and qualitative predator and prey dynamics. Neural networks represent the unknown drift and diffusion functions. The following subsections describe these functions and the fitting objective, summarized in Figure~\ref{fig:model_architecture_training}.

\subsection{Effective stochastic dynamics}\label{sec:effective_stochastic_dynamics}

We seek a two variable description of predator and prey interactions, using population abundances in the biological systems and turbulence and zonal flow energies in the plasma system. These variables do not resolve all the processes that influence their evolution. Consequently, observations starting from similar states can exhibit different subsequent changes. We represent their average evolution and variability through an SDE, written in It\^o form as
\begin{equation}
dX_t=f_\theta(X_t)\,dt+G_\theta(X_t)\,dW_t,
\qquad
Q_\theta(x)=G_\theta(x)G_\theta(x)^\top .
\label{eq:effective_sde}
\end{equation}
Here $X_t=(X_{1,t},X_{2,t})$, with the first variable playing the role of prey and the second the role of predator. The drift $f_\theta$ describes the systematic rate of change, while $G_\theta$ sets the amplitudes and coupling of the random fluctuations. Neural networks represent these state dependent functions, with their adjustable parameters denoted collectively by $\theta$. This allows the interaction law to be inferred without prescribing a particular predator and prey equation.

The components of $dW_t$ are independent Gaussian increments with zero mean and variance $dt$. The matrix $Q_\theta$ gives the fluctuation covariance per unit time, including correlations induced by $G_\theta$. We assume that the current state determines the statistics of subsequent changes (the Markov assumption).

The physical meaning of the coefficients can be expressed through changes over a short interval. For $\Delta X=X_{t+\Delta t}-X_t$, the conditional moments satisfy
\begin{equation}
\begin{aligned}
\mathbb E[\Delta X\mid X_t=x]
 &=f_\theta(x)\Delta t+o(\Delta t),\\
\operatorname{Cov}[\Delta X\mid X_t=x]
 &=Q_\theta(x)\Delta t+o(\Delta t).
\end{aligned}
\label{eq:conditional_increment_moments}
\end{equation}
The average is over realizations starting at $x$. As $\Delta t\to0$, the mean increment per unit time gives the drift, and its covariance per unit time gives $Q_\theta$~\cite{boninsegna2018}. The fitted coefficients describe effective motion and fluctuations at the observation scale.

To fit the model, we use pairs of observed states $x_n$ and $x_{n+1}$, separated by an interval $\Delta t_n$. Holding the drift and diffusion fixed at the starting state gives the Euler--Maruyama approximation: the next state is Gaussian distributed around the drift prediction,
\begin{equation}
x_{n+1}\mid x_n
\;\approx\;
\mathcal N\!\left(
x_n+f_\theta(x_n)\Delta t_n,\,
Q_\theta(x_n)\Delta t_n+\epsilon I
\right).
\label{eq:euler_maruyama_transition}
\end{equation}
Here $\mathcal N$ denotes a Gaussian distribution specified by its mean and covariance, and the small diagonal term $\epsilon I$ makes the covariance numerically invertible.
% It is a numerical regularizer rather than a separately inferred measurement-noise model. The approximation neglects changes in the coefficients during the observation interval and can therefore become less accurate for widely separated observations.

We determine the network parameters by minimizing the negative log likelihood of the observed transitions,
\begin{equation}
\mathcal L_{\mathrm{NLL}}
=\frac12\mathbb E_n\!\left[
r_n^\top\Sigma_n^{-1}r_n
+\log|\Sigma_n|+2\log(2\pi)
\right],
\label{eq:transition_negative_log_likelihood}
\end{equation}
where
\begin{equation}
\begin{aligned}
r_n&=x_{n+1}-x_n-f_\theta(x_n)\Delta t_n,\\
\Sigma_n&=Q_\theta(x_n)\Delta t_n+\epsilon I.
\end{aligned}
\label{eq:transition_residual_covariance}
\end{equation}
Here $\mathbb E_n$ is an average over a fitting batch and $|\Sigma_n|$ is the determinant. The residual $r_n$ measures the departure from the predicted mean motion. The quadratic term penalizes this departure relative to the predicted fluctuations, while $\log|\Sigma_n|$ prevents the fit from reducing that penalty simply by increasing the noise. Drift and diffusion are therefore learned jointly. With few transitions near each state, however, many coefficient fields can generate the same statistics. The additional guidance below helps constrain them.

\subsection{Drift and diffusion terms}\label{sec:drift_diffusion_parameterization}

We represent the systematic evolution as the amount of each
population or energy multiplied by a state dependent growth or
decay rate. This structure is familiar from population growth
and energy balance equations. The neural networks determine how
these rates depend nonlinearly on both variables.

The drift is parameterized as
\begin{equation}
f_i(x)=\widetilde x_i B_f
\tanh\!\left(\frac{h_i(z)}{B_f}\right),
\qquad
\widetilde x_i=\max(x_i,10^{-5}).
\label{eq:bounded_state_scaled_drift}
\end{equation}
Here $i=1,2$, and $h_i$ are outputs of the neural network for drift, evaluated at the standardized state $z$ defined below. Away from the numerical floor, $f_i/x_i$ is the fractional growth or decay rate. For $|h_i|\ll B_f$, $f_i\approx\widetilde x_i h_i$, while the hyperbolic tangent bounds the rate so that $|f_i|\leq B_f\widetilde x_i$. We use $B_f=2$ to limit extreme inferred rates in poorly sampled regions.

The input $z=(x-\mu_{\mathrm{train}})/s_{\mathrm{train}}$ uses componentwise training means and standard deviations to put both variables on comparable numerical scales. Predicted coefficients remain in the original state coordinates. Separate drift and diffusion networks each have two hidden layers with the smooth activation $\operatorname{SiLU}(u)=u/(1+e^{-u})$.

For the stochastic contribution, we separate the fluctuation
amplitude of each variable from the correlation between them:
\begin{equation}
Q_\theta(x)=
\begin{pmatrix}
s_1(x)^2 & \rho(x)s_1(x)s_2(x)\\
\rho(x)s_1(x)s_2(x) & s_2(x)^2
\end{pmatrix},
\label{eq:diffusion_covariance}
\end{equation}
where
\begin{equation}
\begin{aligned}
s_i(x)
 &=B_\sigma\operatorname{sigmoid}(a_i(z))
   \widetilde x_i+10^{-5},\\
\rho(x)&=\tanh(r(z)).
\end{aligned}
\label{eq:noise_amplitudes_correlation}
\end{equation}
Here $a_1,a_2,r$ are outputs of the diffusion network,
$\operatorname{sigmoid}(u)=1/(1+e^{-u})$, and
$B_\sigma=0.35$. The amplitude $s_i$ gives a stochastic
increment with variance $s_i^2\,dt$ over a short interval $dt$.

We assume multiplicative fluctuations, with the sigmoid bounding their relative amplitude. Away from the numerical floors, the rms increment relative to $x_i$ is $B_\sigma\operatorname{sigmoid}(a_i(z))\sqrt{dt}$. The correlation $\rho$ allows fluctuations of the same or opposite sign after subtracting the drift. %It describes instantaneous noise coupling, rather than the correlation of the full time series, which also depends on the deterministic interaction.

The transformations in Eq.~\eqref{eq:noise_amplitudes_correlation} ensure a valid
covariance: $s_i>0$, $|\rho|\leq1$, and
\begin{equation}
\det Q_\theta=s_1^2s_2^2(1-\rho^2)\geq0.
\label{eq:diffusion_covariance_determinant}
\end{equation}
Together with the nonnegative diagonal entries, this guarantees
nonnegative variance in every direction in state space.

A compatible matrix for generating the stochastic increments is
\begin{equation}
G_\theta=
\begin{pmatrix}
s_1&0\\
\rho s_2&s_2\sqrt{1-\rho^2}
\end{pmatrix},
\qquad
G_\theta G_\theta^\top=Q_\theta.
\label{eq:diffusion_cholesky_factor}
\end{equation}
The first variable receives one noise, and the second receives a mixture of that shared noise and an independent noise. 
%This produces the required covariance without implying a direction of physical influence. 
Any factor with the same product $GG^\top=Q_\theta$ gives the same local Gaussian fluctuation statistics.

The coefficient formulas include numerical floors near zero. During simulation, each state component is bounded below by $10^{-6}$ after every integration update. Appendix~\ref{sec:training_optimization_details} gives the numerical details.

\subsection{Adaptive local moment guidance}\label{sec:adaptive_local_moment_guidance}

To separate systematic motion from fluctuations, we pool transitions starting near a state of interest.
Let $\Delta x_n=x_{n+1}-x_n$ be an observed increment, and let
$q$ be a state selected from the training observations at which
we wish to estimate the coefficients. Within its neighborhood,
we approximate the mean increment by a linear function:
\begin{equation}
\Delta x_n\approx b_0+B(x_n-q),
\qquad
\widehat f(q)=\frac{b_0}{\Delta t_{\mathrm{ref}}}.
\label{eq:local_linear_drift_estimate}
\end{equation}
The intercept $b_0$ estimates the mean increment at $q$, and $B$ accounts for its variation across nearby starting states. Dividing by the median training interval $\Delta t_{\mathrm{ref}}$ gives a local drift estimate.

The intercept and slopes are obtained by minimizing
\begin{equation}
\sum_n w_n\|\Delta x_n-b_0-B(x_n-q)\|_2^2
+\eta\left(\|b_0\|_2^2+\|B\|_F^2\right).
\label{eq:local_linear_regression_objective}
\end{equation}
The weights $w_n=w_n(q)$ follow a Gaussian profile within the neighborhood, vanish outside it, and sum to approximately one. Distances are measured in units of each variable's training standard deviation. The small penalty $\eta=10^{-6}$ regularizes poorly determined intercepts and slopes. Here $\|\cdot\|_2$ and $\|\cdot\|_F$ denote the Euclidean vector norm and the Frobenius matrix norm. Appendix~\ref{sec:training_optimization_details} gives the weighting details.

After subtracting the fitted local mean, the residual
$e_n=\Delta x_n-b_0-B(x_n-q)$ measures the unexplained part
of each increment. Its weighted second moment supplies the
local diffusion target:
\begin{equation}
\widehat Q(q)=\frac{1}{\Delta t_{\mathrm{ref}}}
\sum_n w_n e_ne_n^\top .
\label{eq:local_covariance_estimate}
\end{equation}
Dividing the residual second moment by $\Delta t_{\mathrm{ref}}$ estimates covariance per unit time. The matrix is symmetrized and its eigenvalues are given a small positive floor for numerical stability (Appendix~\ref{sec:training_optimization_details}).
For unequal observation intervals, division by a common $\Delta t_{\mathrm{ref}}$ makes these moment targets approximate.

Neighborhoods expand where data are sparse and contract where they are dense, balancing statistical reliability against resolution of state dependence. Validation selects the neighborhood settings. Appendix~\ref{sec:training_optimization_details} specifies them for each application.

% These estimates remain approximate. Local linear fitting
% reduces the contribution of systematic spatial variation to
% the residuals, but cannot remove all effects of nonlinear
% variation, finite observation intervals, or measurement noise.
% For irregularly sampled records, using a common
% $\Delta t_{\mathrm{ref}}$ introduces a further approximation:
% increments measured over different durations are pooled
% without explicitly accounting for those durations in the
% local regression. The transition likelihood in
% Section~\ref{sec:effective_stochastic_dynamics}, in contrast, uses each recorded
% $\Delta t_n$.

The local estimates guide the neural coefficients through
two additional fitting penalties in the form of mean-squared errors, given by
\begin{equation}
\begin{aligned}
\mathcal L_{\mathrm{moment},f}
 &=\frac{\mathbb E_q
 \|f_\theta(q)-\widehat f(q)\|_2^2}{2c_f},\\
\mathcal L_{\mathrm{moment},Q}
 &=\frac{\mathbb E_q
 \|Q_\theta(q)-\widehat Q(q)\|_F^2}{4c_Q}.
\end{aligned}
\label{eq:moment_matching_losses}
\end{equation}
Here $\mathbb E_q$ averages over sampled training states. The penalties match the neural drift and covariance to the local estimates, with factors 2 and 4 averaging over their components.

For the biological and HW applications, the normalization
scales are
\begin{equation}
\begin{aligned}
c_f&=\max\!\left\{
\tfrac12\mathbb E_q\|\widehat f(q)\|_2^2,\,10^{-4}
\right\},\\
c_Q&=\max\!\left\{
\tfrac14\mathbb E_q\|\widehat Q(q)\|_F^2,\,10^{-6}
\right\}.
\end{aligned}
\label{eq:moment_normalization_scales}
\end{equation}
The scales use the same query batch and express errors relative to the target magnitudes. Their floors avoid division by zero. The targets remain fixed during neural fitting. %

\subsection{Predator and prey phase space guidance}\label{sec:phase_space_guidance}

Local increment statistics constrain the dynamics where
observations are available, but a short record may leave much
of the state space poorly sampled. We supplement this
information with two qualitative expectations for the
deterministic drift. Near extinction, prey can grow while
predators decline in the absence of sufficient prey. Near
coexistence, small disturbances are expected to decay under
the deterministic dynamics. In the turbulence and zonal flow
interpretation, these expectations correspond to turbulence
growth when flow suppression is weak, flow damping when its
turbulent drive is absent, and a restoring tendency near a
state where both energies are finite.
These are modeling assumptions for the systems considered, rather than universal properties of predator and prey dynamics. We encode them as soft penalties in the fitting objective.

We express these preferences through the drift Jacobian $J_f(x)=\partial f_\theta/\partial x$. Its eigenvalues describe the growth or decay of small perturbations at an equilibrium.

We first consider a state near zero density
$x^0=(10^{-5},10^{-5})$ and write $J_0=J_f(x^0)$.
At the state floor, derivatives are taken from the side $x_i>10^{-5}$.
For an ideal multiplicative predator and prey drift near the
origin, positive prey growth and predator decay correspond
to $(J_0)_{11}>0$ and $(J_0)_{22}<0$.
We also favor $\det J_0<0$, which gives one positive and one
negative eigenvalue: the local signature of a saddle.
The penalty is
\begin{equation}
\mathcal L_{\mathrm{saddle}}
=\psi\!\left(\delta-(J_0)_{11}\right)
+\psi\!\left(\delta+(J_0)_{22}\right)
+\psi\!\left(\delta+\det J_0\right),
\label{eq:extinction_saddle_penalty}
\end{equation}
where $\psi(u)=[\log(1+e^u)]^2$ is a smooth increasing
penalty and $\delta>0$ is a small margin. The three terms
favor positive prey growth, negative predator growth, and
opposite signs of the two eigenvalues, respectively.

% This constraint is applied close to the origin because the
% coefficient floors modify the exact extinction boundary.
% Consequently, it guides the local derivatives of the drift;
% it does not establish an exact saddle equilibrium at $x^0$
% or an absorbing extinction state.

For coexistence, we select a positive reference state
$x^\star$ near a peak of the training state density and hold
it fixed during fitting. Treating this frequently visited state as a stable drift equilibrium is a modeling assumption.

Two penalties encourage a restoring drift near this anchor:
\begin{equation}
\begin{aligned}
\mathcal L_{\mathrm{fp}}
 &=\|f_\theta(x^\star)\|_2^2,\\
\mathcal L_{\mathrm{stable}}
 &=\psi\!\left(\delta+\operatorname{tr}J_\star\right)
  +\psi\!\left(\delta-\det J_\star\right),
\qquad
J_\star=J_f(x^\star).
\end{aligned}
\label{eq:coexistence_equilibrium_stability_penalties}
\end{equation}
The first penalty favors zero drift at the anchor. The second favors a negative trace and positive determinant. At an equilibrium in two dimensions, these signs imply eigenvalues with negative real parts. Both a stable node and a stable focus are allowed. A focus permits damped oscillations that stochastic forcing can repeatedly excite.

The combined contribution to the fitting objective is,
up to terms independent of the network parameters,
\begin{equation}
\mathcal L_{\mathrm{topology}}(e)
=\alpha(e)\left(
\lambda_{\mathrm{s}}\mathcal L_{\mathrm{saddle}}
+\lambda_{\mathrm{fp}}\mathcal L_{\mathrm{fp}}
+\lambda_{\mathrm{st}}\mathcal L_{\mathrm{stable}}
\right).
\label{eq:topology_guidance_loss}
\end{equation}
The weights $\lambda_{\mathrm{s}},\lambda_{\mathrm{fp}},\lambda_{\mathrm{st}}$ set the relative strengths of the constraints. The factor $\alpha(e)$, a function of the training epoch $e$, rises from zero to one after an initial training period, allowing the networks to fit the observations before applying full dynamical guidance. Appendix~\ref{sec:training_optimization_details} gives the weights and schedule.

Here, ``topology constrained'' denotes soft guidance of the local drift near extinction and coexistence. It does not guarantee the global phase space topology of the fitted model.

\subsection{Joint objective and coefficient regularization}
\label{sec:joint_objective_regularization}

Sparse observations can admit several plausible coefficient fields. We add regularization that favors smaller drift and noise amplitudes and weaker noise correlation when the other fitting terms do not distinguish them.

For the biological and HW applications, this contribution is
\begin{equation}
\begin{aligned}
\mathcal L_{\mathrm{reg}}
={}&\lambda_f
\frac{\mathbb E_n\|f_\theta(x_n)\|_2^2}{2d_f}
+\lambda_Q
\frac{\mathbb E_n\operatorname{tr}Q_\theta(x_n)}{d_Q}\\
&+\lambda_\rho\mathbb E_n\rho_\theta(x_n)^2.
\end{aligned}
\label{eq:coefficient_regularization}
\end{equation}
Here $\mathbb E_n$ denotes an average over a batch of training
transitions, and the nonnegative weights $\lambda_f$,
$\lambda_Q$, and $\lambda_\rho$ control the three penalties.
The correlation $\rho_\theta$ is defined in
Section~\ref{sec:drift_diffusion_parameterization}.

% The first term discourages large systematic rates of change.
% The second penalizes
% $\operatorname{tr}Q_\theta=s_1^2+s_2^2$, the sum of the
% instantaneous fluctuation variances in the two state
% coordinates. It supplements the likelihood, which already
% balances prediction residuals against the covariance assigned
% to them. The third term favors noise correlation closer to
% zero, while allowing either sign when supported by the
% other fitting terms. It acts on the instantaneous stochastic
% coupling, not on correlations generated by the deterministic
% interaction between the variables.

To relate the amplitude penalties to the scale of the
observations, we use
\begin{equation}
\begin{aligned}
d_f&=\max\!\left\{
\tfrac12\mathbb E_n
\|\Delta x_n/\Delta t_n\|_2^2,\,10^{-6}
\right\},\\
d_Q&=\max\!\left\{
\mathbb E_n\left[
\|\Delta x_n\|_2^2/\Delta t_n
\right],\,10^{-6}
\right\},
\end{aligned}
\label{eq:regularization_normalization_scales}
\end{equation}
where $\Delta x_n=x_{n+1}-x_n$.
The first scale measures the mean squared observed rate of
change, averaged over the two components. The second measures
the mean squared increment per unit time. These quantities
provide reference magnitudes for the penalties.
%  they are not
% separate estimates of the true drift and diffusion, since
% each observed increment contains both contributions.
% They are computed from the current training batch and do
% not depend on the network parameters. 
The positive lower
bounds prevent division by nearly zero values.

% These penalties express a preference during estimation.
% They do not imply that the physical drift, fluctuations, or
% noise correlation should vanish. Their influence is balanced
% against the likelihood, moment targets, and dynamical guidance.
% Likewise, penalizing coefficient magnitudes at observed states
% does not guarantee stability elsewhere in state space.

The complete fitting objective is
\begin{equation}
\begin{aligned}
\mathcal L={}&
\mathcal L_{\mathrm{moment},f}
+\mathcal L_{\mathrm{moment},Q}
+\lambda_{\mathrm{NLL}}\mathcal L_{\mathrm{NLL}}\\
&+\mathcal L_{\mathrm{topology}}
+\mathcal L_{\mathrm{reg}},
\end{aligned}
\label{eq:joint_training_objective}
\end{equation}
with $\lambda_{\mathrm{NLL}}=0.5$ and unit weights for the
two normalized moment penalties. Minimizing this objective
seeks a compromise: the model should explain individual
observed changes, reproduce their local mean and covariance,
retain the preferred local drift structure, and avoid
unnecessarily large coefficients.

Training uses AdamW, a gradient-based optimizer with decoupled weight decay, together with gradient clipping to limit abrupt updates. Validation selects the retained model. Each network has two hidden layers, with 32 units per layer for biology and 64 for HW. Appendix~\ref{sec:training_optimization_details} gives the optimization settings, penalty weights, and differences used for the controlled reference.

\section{Assessments and comparison models}\label{sec:assessment_comparison_models}

We assess whether the learned models describe both systematic evolution and stochastic variability when fitting observations are limited. Three types of experiment provide complementary tests. \revadd{A prescribed nonlinear stochastic equation provides a controlled test of learning from limited fitting transitions.} HW simulations test a two energy description of spatial plasma dynamics, with independent trajectories for evaluation. Biological records test observed population changes from short measured records. In both applications, the generating two variable coefficients are unknown. Sections~\ref{sec:known_equation_controls} to \ref{sec:biological_applications} introduce each system and its results. Here we define the assessment criteria and comparison models.

\subsection{Assessment criteria}\label{sec:assessment_criteria}

A stochastic model specifies a distribution of possible changes from a given state. Its assessment therefore requires more than agreement between a predicted trajectory and a measured one. We examine the accuracy of predicted changes and the statistical behavior of repeated simulations. \revadd{The comparisons use conditional-mean prediction, transition distributions, and, where independent trajectories are available, autonomously simulated state distributions.}

\subsubsection{Normalized root mean square error}\label{sec:prediction_nrmse}

Repeated simulations from each observed starting state estimate the mean endpoint. We compare this prediction with the measured endpoint using normalized root mean square error (NRMSE), given by
\begin{equation}
\operatorname{NRMSE}=\sqrt{\frac{1}{2N}\sum_n
\left\|\frac{\widehat\mu_n-x_{n+1}}{s_{\mathrm{ref}}}\right\|_2^2},
\qquad \widehat\mu_n=\frac{1}{64}\sum_{r=1}^{64}\widetilde x_{n+1}^{(r)}.
\label{eq:conditional_mean_nrmse}
\end{equation}

Here $N$ is the number of test transitions and $\widetilde x_{n+1}^{(r)}$ is the endpoint of realization $r$ starting at the observed state $x_n$. The vector $s_{\mathrm{ref}}$ contains reference standard deviations used to put the two variables on comparable scales.

\revadd{For biological records and HW energies, each realization is integrated over one recorded interval $\Delta t_n$ using four Euler--Maruyama substeps of duration $\delta t_n=\Delta t_n/4$. The nonlinear control uses a fixed numerical step of $0.01$, giving 25 and 50 substeps for observation intervals $0.25$ and $0.5$, respectively (Appendix~\ref{sec:nonlinear_control_simulation}).} 
The 64 realizations are repeated predictions from the same observed starting state. 
Their endpoints are averaged before computing NRMSE. 
Every model uses the same integration procedure and random streams within a comparison, except NeuralODE, which has zero diffusion. 
%HW additionally tests drift only rollouts from 128 test windows. 
% Starting at each observed initial state, we integrate $\dot x=f_\theta(x)$ with no sampled noise over four observation intervals, using four Euler substeps per interval and no resetting to subsequent observations. The rollout error compares these deterministic predictions with the observed trajectories. For nonlinear drift, such rollouts generally differ from the SDE conditional mean, so this error includes stochastic variability as well as prediction error. Here four observation intervals specify the forecast horizon; with four substeps per interval, the forecast uses 16 numerical updates.

Each measured endpoint contains fluctuations around its conditional mean, so prediction NRMSE includes both model error and stochastic variability. It need not vanish even when the model is correctly specified.
\textbf{NRMSE primarily assesses mean prediction, which is closely related to drift.}

\subsubsection{Maximum mean discrepancy}\label{sec:maximum_mean_discrepancy}

Squared maximum mean discrepancy (MMD$^2$) measures agreement between two distributions through similarities among their samples \cite{gretton2012}. Let $u_1,\ldots,u_N$ be reference samples and $v_1,\ldots,v_M$ be simulated samples, expressed in the same standardized coordinates. The empirical estimate is
\begin{equation}
\begin{aligned}
\widehat{\operatorname{MMD}}^{\,2}
={}&\frac{1}{N^2}\sum_{i,j=1}^{N}k(u_i,u_j)
 +\frac{1}{M^2}\sum_{i,j=1}^{M}k(v_i,v_j)\\
 &-\frac{2}{NM}\sum_{i=1}^{N}\sum_{j=1}^{M}k(u_i,v_j).
\end{aligned}
\label{eq:mmd_empirical}
\end{equation}
The kernel $k$ measures similarity. We use an average of Gaussian kernels,
\begin{equation}
k(u,v)=\frac{1}{L}\sum_{\ell=1}^{L}
\exp\!\left[-\frac{\|u-v\|_2^2}{2h_\ell^2}\right],
\label{eq:mmd_kernel}
\end{equation}
where $h_\ell>0$ are bandwidths that set the distances over which samples are considered similar. Combining bandwidths compares the distributions at several scales.

The first two terms in Eq.~\eqref{eq:mmd_empirical} measure similarity within each sample set. The last measures similarity between the sets. The estimate is small when simulated and reference samples occupy similar regions with similar relative frequencies. The sums include pairs with $i=j$, giving the biased, nonnegative empirical estimate. Lower values indicate closer agreement. This comparison is sensitive to spread and dependence as well as the mean.

We use two forms of this metric, distinguished by the samples supplied to Eq.~\eqref{eq:mmd_empirical}.

\textit{Transition MMD$^2$} compares observed and simulated pairs of starting and ending states:
\begin{equation}
\begin{aligned}
u_n&=(x_n,x_{n+1}),\\
v_n^{(r)}&=(x_n,\widetilde x_{n+1}^{(r)}).
\end{aligned}
\label{eq:transition_mmd_samples}
\end{equation}
Each sample has four coordinates. The simulated and observed pairs share the same starting state, and each simulated endpoint is obtained with the simulation protocol of Section~\ref{sec:prediction_nrmse}. We compute MMD$^2$ separately for each of the 64 endpoint sets and average the scores, rather than averaging the endpoints first. 
\textbf{Transition MMD$^2$ assesses the joint distribution of starting and ending states, testing the combined effects of drift and diffusion over the observation interval.}

% In the known-equation controls of Section~\ref{sec:known_equation_controls}, transition MMD$^2$ uses all 225 held-out pairs per seed. Both states in each pair are standardized using the training-start means and standard deviations, retaining the $10^{-4}$ addition to the latter in the nonlinear control. The three Gaussian bandwidths are 0.5, 1, and 2 times the median positive distance between distinct standardized four-dimensional training pairs. These settings are shared across models within each seed.

% We use it for both biological records and HW energies, which directly supply observed transitions. Restarting at each observed state tests the interaction over the measured intervals without requiring the record to establish a stationary state distribution. The score summarizes agreement across all sampled starts. In HW, this local test complements state MMD$^2$: agreement over one observed interval need not ensure that autonomous simulations visit the correct energy ranges.

\textit{State MMD$^2$} compares individual states visited by reference and learned trajectories:
\begin{equation}
u_i=x(t_i),\qquad v_j=\widetilde x(s_j).
\label{eq:state_mmd_samples}
\end{equation}
Each sample has two coordinates. Learned trajectories start from observed states and then evolve autonomously with drift and, for stochastic models, noise. Samples collected after an initial discarded period are pooled for comparison. 
\textbf{State MMD$^2$ measures which regions of state space are visited and their relative frequencies over the evaluated duration.} It compares occupancy rather than the temporal order of the states.

Together, these metrics assess mean prediction, transition statistics, and autonomous state distributions, providing complementary tests of the learned stochastic dynamics.

% We use state MMD$^2$ for both known systems and HW, where simulated reference trajectories allow this test of autonomous behavior. In the controls, the reference trajectories come from the prescribed SDEs. In HW, they are energy trajectories from independent field simulations. This asks whether the learned two variable model reproduces the distribution generated by the full spatial system. Transition and state MMD$^2$ answer different questions, so their values are compared within each experiment.

% The integration steps for these stochastic trajectories differ from the one step drift prediction. The gLV control uses eight paths over eight time units with step $0.03$. The nonlinear control uses 12 paths over eight time units with step $0.0125$, one quarter of its observation interval. HW uses 12 paths over 768 observation intervals with four substeps per interval, giving step $0.25$. Completion criteria, discarded initial periods, sample selection, scaling, and bandwidths are given in Appendices~\ref{sec:biological_validation_details}, \ref{sec:nonlinear_control_simulation}, and \ref{sec:hw_energy_model_details}. MMD$^2$ values are compared within each experiment.

\subsubsection{\revadd{Relative errors in drift and covariance}}\label{sec:relative_coefficient_rmse}

\revadd{For the known-equation test, we compare the learned drift $f_\theta$ and diffusion covariance $Q_\theta$ with the generating coefficients $f$ and $Q=GG^\top$ at the test starting states $x_n$. Their relative root mean square errors (rRMSE) are}
{\ifshowrevision\color{black}\fi
\begin{equation}
\begin{aligned}
\operatorname{rRMSE}_f
&=\left[\frac{\sum_n\|f_\theta(x_n)-f(x_n)\|_2^2}
{\sum_n\|f(x_n)\|_2^2}\right]^{1/2},\\
\operatorname{rRMSE}_Q
&=\left[\frac{\sum_n\|Q_\theta(x_n)-Q(x_n)\|_F^2}
{\sum_n\|Q(x_n)\|_F^2}\right]^{1/2}.
\end{aligned}
\label{eq:relative_coefficient_rmse}
\end{equation}}
\revadd{The squared vector norm sums over the drift components, and the squared Frobenius norm sums over all covariance entries. Lower errors indicate closer agreement with the generating coefficients. For NeuralODE, $Q_\theta=0$, so $\operatorname{rRMSE}_Q=1$.}

\subsection{Comparison models}\label{sec:comparison_models}

The main comparisons use five models: stochastic gLV, NeuralODE, positive neural SDE, DynGMA, and TC-AdaNSDE. They examine whether a conventional interaction law is sufficient, what is gained by flexible stochastic coefficients, and how alternative fitting procedures perform. Within each experiment, the evaluated models use the same fitting observations. Table~\ref{tab:main_model_comparison} summarizes their differences. Additional structural comparisons and matched guidance ablations are given in Appendix~\ref{sec:hw_guidance_ablation}.

The stochastic generalized Lotka--Volterra (gLV) model represents a conventional interaction law with multiplicative noise:

\begin{equation}
\begin{aligned}
f_i^{\mathrm{gLV}}(x)&=\widetilde x_i\left(a_i+\sum_{j=1}^2 b_{ij}x_j\right),\\
Q_{ii}^{\mathrm{gLV}}(x)&=\sigma_i^2\widetilde x_i^2,\qquad
Q_{12}^{\mathrm{gLV}}(x)=\rho\sigma_1\sigma_2\widetilde x_1\widetilde x_2.
\end{aligned}
\label{eq:stochastic_glv_coefficients}
\end{equation}

The parameters $a_i,b_{ij}$ determine growth and interactions, $\sigma_i>0$ are constant fractional noise amplitudes, and $\rho$ is a constant noise correlation. The state floor $\widetilde x_i$ is defined in Section~\ref{sec:drift_diffusion_parameterization}. All nine parameters are fitted through the Gaussian transition likelihood, with coefficient regularization in the applications and model selection using validation data. This reference tests whether a compact classical equation already describes the observations.

The neural ordinary differential equation (NeuralODE) learns a nonlinear drift by minimizing squared prediction error, with $Q=0$. It tests the description available from deterministic nonlinear dynamics.

The positive neural SDE uses state-scaled drift, softplus fractional noise amplitudes, and a learned noise correlation. It retains numerical floors but omits TC-AdaNSDE's drift and diffusion saturation bounds. It is fitted through the Gaussian transition likelihood without local moment guidance or topology constraints.

The dynamical Gaussian mixture approximation (DynGMA) of Zhu and Li \cite{zhu2024} addresses how the transition distribution is approximated over a finite observation interval. It propagates a mixture of Gaussian components through intermediate states to account for variation in the coefficients. We apply its likelihood to the same positive neural drift and full covariance architecture as the positive neural SDE. It therefore tests whether improving the transition approximation provides an effective alternative to the statistical and dynamical guidance used here. Appendix~\ref{sec:training_optimization_details} gives the implementation based on the authors' Algorithm 2 for variable observation intervals.

TC-AdaNSDE combines state scaling and correlated diffusion with the adaptive local moment guidance and phase space penalties developed in Section~\ref{sec:methodology}. The latter favor the prescribed behavior near extinction and coexistence. It also bounds the fractional drift and noise amplitudes using the saturation functions in Section~\ref{sec:drift_diffusion_parameterization}. The comparisons assess the performance of this combined construction relative to the other four principal models. Appendix~\ref{sec:training_optimization_details} gives the fitting details.

% No extra packages required in revtex4-2

% Add to your preamble:
% \usepackage{pifont}
% \newcommand{\cmark}{\ding{51}}
% \newcommand{\xmark}{\ding{55}}

\begin{table*}[htbp]
\caption{The five principal comparison models. The models differ in their interaction law, representation of fluctuations, and fitting objective. \revadd{Coefficient regularization is used for the stochastic models in the application fits and nonlinear control.} Additional structural comparisons and matched guidance ablations are given in Appendix~\ref{sec:hw_guidance_ablation}.}
\label{tab:main_model_comparison}
\fontsize{9.5}{11.5}\selectfont
\setlength{\tabcolsep}{4pt}
\renewcommand{\arraystretch}{1.18}
\begin{tabularx}{\linewidth}{@{}>{\hsize=0.85\hsize\raggedright\arraybackslash}X>{\hsize=1.1\hsize\raggedright\arraybackslash}X>{\hsize=1.1\hsize\raggedright\arraybackslash}X>{\hsize=1.15\hsize\raggedright\arraybackslash}X>{\hsize=0.8\hsize\raggedright\arraybackslash}X@{}}
\toprule
\textbf{Model} & \textbf{Drift} & \textbf{Diffusion} & \textbf{Data-fitting objective} & \textbf{Additional guidance} \\
\midrule
Stochastic gLV & State-scaled affine interaction law & Constant fractional amplitudes and noise correlation & Gaussian transition likelihood & None \\
NeuralODE & Neural drift & None ($Q=0$) & Squared prediction error & None \\
Positive neural SDE & State-scaled neural drift without saturation bounds & State-scaled softplus amplitudes; learned state-dependent correlation & Gaussian transition likelihood & None \\
DynGMA neural SDE & Same as positive neural SDE & Same as positive neural SDE & Gaussian mixture transition likelihood & None \\
TC-AdaNSDE & State-scaled neural drift with bounded fractional rates & State-scaled, bounded neural amplitudes; learned state-dependent correlation & Gaussian transition likelihood & Adaptive local moments; extinction and coexistence penalties \\
\bottomrule
\end{tabularx}
\end{table*}

% \subsection{Comparisons with limited observations}\label{sec:limited_observation_comparisons}

% Within each experiment, models share the fitting observations and the validation and test sets. Training observations determine the coefficient normalization, local moment targets, and coexistence anchor. Validation selects neighborhood settings and fitted checkpoints. The test observations are used to calculate the reported errors. Repeated fits measure variation due to initialization and, where specified, the training trajectories.

% The biological and HW experiments vary the amount of fitting data using nested training sets, with validation and test observations held fixed. Biological records are divided into temporal blocks, whereas HW uses separate complete trajectories. The biological tests assess transitions among comparable observed population states. The HW tests assess new field realizations. Reported fitting counts exclude the additional validation observations. The sections devoted to each system give these counts and explain how observations are selected.

% The prescribed equations provide an initial check of coefficient reconstruction and simulated behavior. The applications then test whether useful prediction and stochastic statistics can be obtained when the two variable interaction law is unknown. This distinction guides the interpretation of the results below.

\section{\revnew{Known nonlinear dynamics and justification of evaluation metrics}}\label{sec:known_equation_controls}

\revnew{In empirical physical and biological systems, the true drift and diffusion coefficients for the two observed variables are unknown. When observations project a system of higher dimension onto two variables, there are no exact generating equations for those variables. Consequently, models learned from real records must be evaluated through their observable behavior rather than direct coefficient errors. Before turning to the plasma and biological applications, we use a nonlinear SDE for predator and prey with known coefficients to examine how conditional mean NRMSE, transition MMD$^2$, and autonomous state MMD$^2$ (Section~\ref{sec:assessment_criteria}) relate to coefficient recovery and why all three are needed when data are limited.}

\subsection{\revadd{Nonlinear interaction and training data}}\label{sec:nonlinear_interaction_control}

\revadd{The generating drift and diffusion are}
\begin{equation}
\begin{gathered}
f(x)=
\begin{pmatrix}
1.2x_1-\dfrac{x_1x_2}{1+0.1x_2^2}\\[6pt]
-0.8x_2+\dfrac{x_1x_2}{1+0.1x_2^2}
\end{pmatrix},\\[3pt]
G(x)=
\begin{pmatrix}
0.20x_1 & 0\\
0.18\rho x_2 & 0.18\sqrt{1-\rho^2}\,x_2
\end{pmatrix},\qquad \rho=-0.60.
\end{gathered}
\label{eq:nonlinear_control_sde}
\end{equation}

The shared transfer term retains a direct coupling between prey loss and predator gain. Its rational dependence on $x_2$ introduces curvature that an affine gLV law for the per capita rates cannot represent exactly, while preserving positive state scaling. \revnew{The covariance remains representable by stochastic gLV, so this system tests learning a nonlinear interaction beyond the prescribed gLV drift family while holding the fluctuation structure fixed.} At these parameters, the deterministic drift has a stable coexistence equilibrium near $(0.969,1.454)$ \revnew{and a second positive saddle equilibrium near $(4.587,6.880)$, making stable coexistence a local property of the sampled regime.}

\begin{figure*}[!t]
\centering
\includegraphics[width=\linewidth]{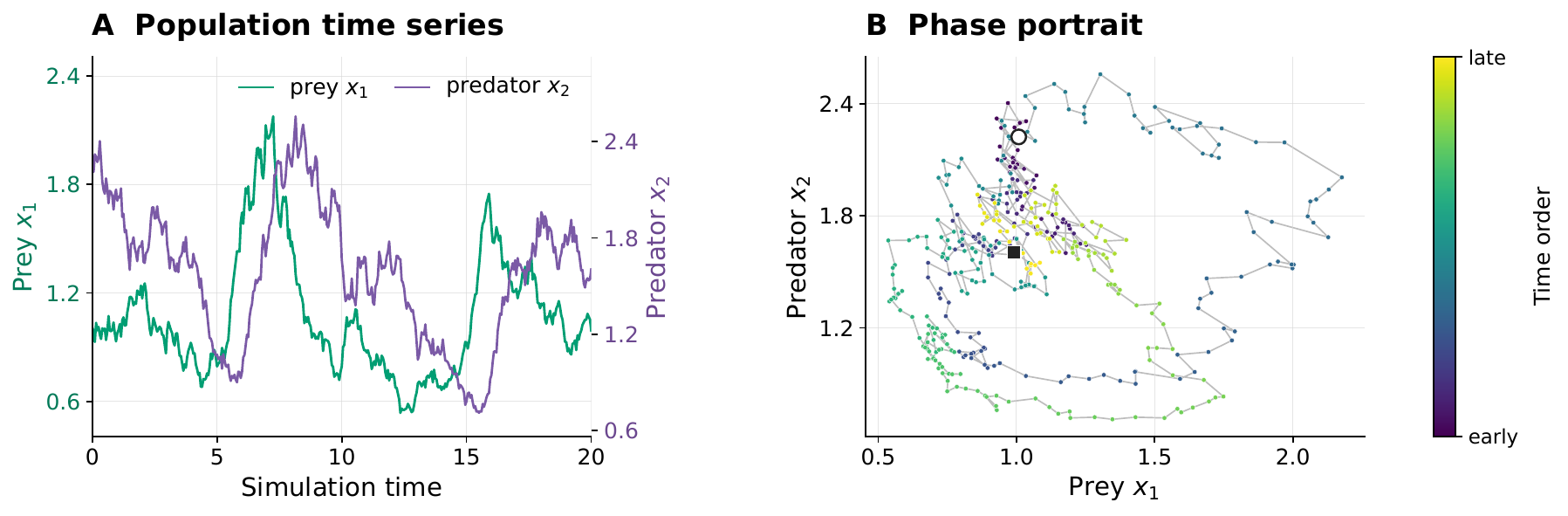}
{\ifshowrevision\color{black}\fi
\caption{\label{fig:known_training_trajectories}A short excerpt from a simulated trajectory of the nonlinear predator and prey system in Eq.~\eqref{eq:nonlinear_control_sde}. (A) Prey and predator populations over the first 20 time units after the initial discarded period. (B) The same segment in the phase plane, with color progressing from early (purple) to late (yellow). The open circle marks the first state and the black square the last.}}
\end{figure*}

\revnew{Figure~\ref{fig:known_training_trajectories} shows a short segment of one simulated trajectory as a time series and a phase portrait. Successive oscillations vary in amplitude and timing, providing a controlled example of irregular predator and prey motion under a known interaction law. We evaluate the models at $\Delta t=0.25$ and at a wider sampling interval $\Delta t=0.5$ to test how sampling resolution affects learning.}

\revnew{For each of three independent repeats, we generate 30 trajectories with numerical step $0.005$, discard the first three time units, and retain 120 time units. Separate groups of five, five, and 20 trajectories supply fitting, validation, and test pairs. The training sets are nested random samples of 100, 300, or 1,050 pairs, evaluated against the same 225 validation and 225 test pairs at every training size and across both observation intervals. Increasing the training budget adds transition pairs from the same histories while keeping the total record duration fixed. Fitting uses each subset's normalization, whereas evaluation scales and kernels remain fixed from the subset of 100 pairs (Appendix~\ref{sec:nonlinear_control_simulation}).}

\subsection{\revnew{Why complementary behavioral metrics are required}}\label{sec:known_equation_interpretation}

\begin{figure*}[!t]
\centering
\includegraphics[width=0.96\linewidth]{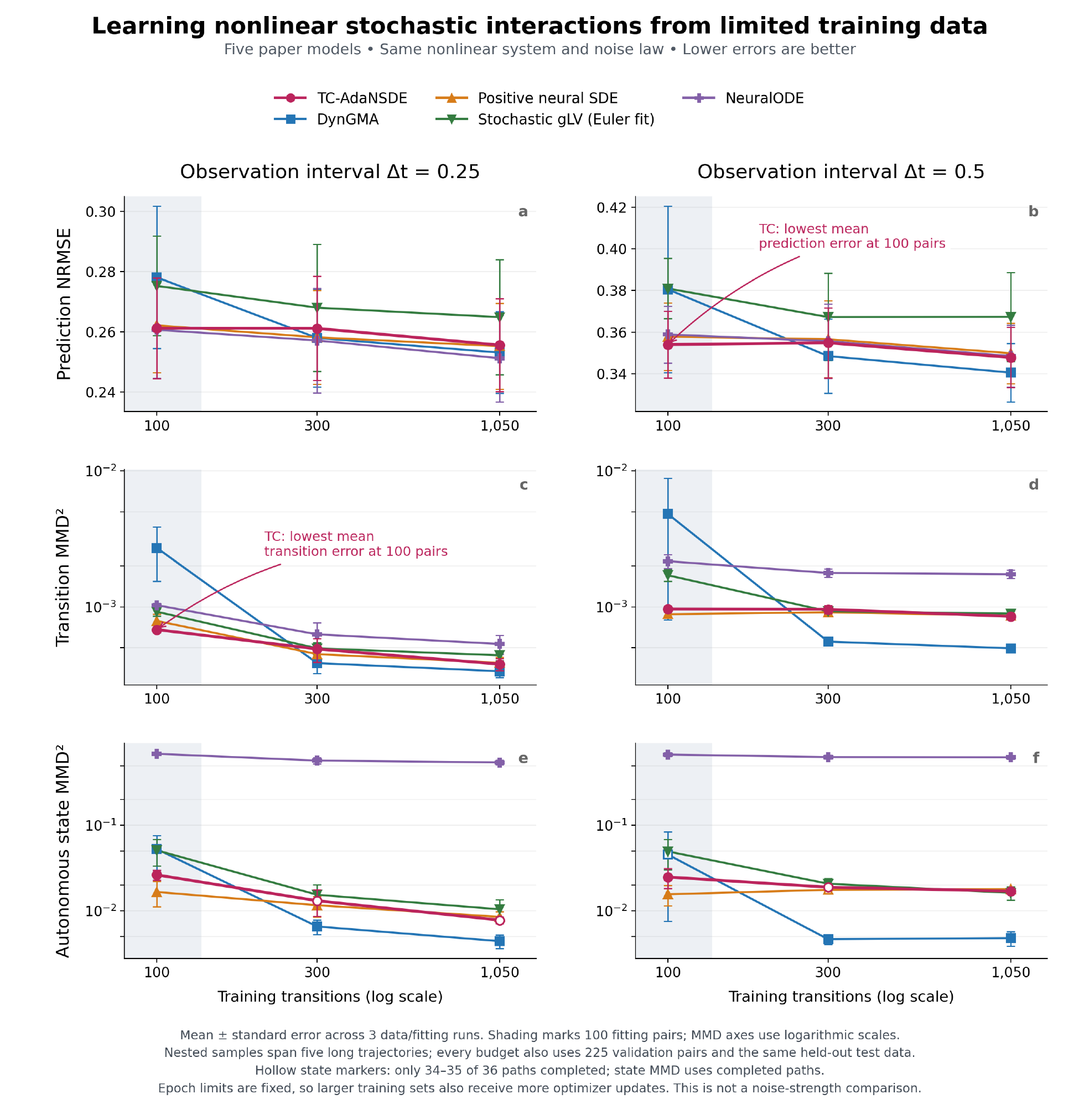}
{\ifshowrevision\color{black}\fi
\caption{\label{fig:known_nonlinear_data_sizes}Learning the nonlinear stochastic interaction from limited fitting data. Columns use observation intervals $0.25$ and $0.5$; rows show conditional mean NRMSE, transition MMD$^2$, and autonomous state MMD$^2$. Shading identifies 100 fitting pairs. Points and error bars are means and one standard error across three data and fitting seeds. Each training size also uses 225 validation pairs and the same test observations. Evaluation scales and kernels are fixed across training sizes within each interval and seed. 
%At settings with hollow state markers, the scores use only the 34 or 35 paths that complete out of 36 attempts. All 36 paths complete elsewhere. Epoch limits are fixed, so larger fitting sets also receive more optimizer updates. The noise law is unchanged.
}}
\end{figure*}

\revnew{Comparing the direct coefficient errors ($\operatorname{rRMSE}_f$ and $\operatorname{rRMSE}_Q$ in Table~\ref{tab:nonlinear_coefficient_errors}) against the behavioral scores in Figure~\ref{fig:known_nonlinear_data_sizes} reveals three key properties of the evaluation metrics.}

\revnew{First, conditional mean NRMSE reliably tracks local drift accuracy across sampling intervals, but cannot by itself distinguish how well a model captures stochastic fluctuations. At 100 training pairs, TC-AdaNSDE achieves low drift error ($\operatorname{rRMSE}_f = 0.3125$ at $\Delta t=0.25$ and $0.3115$ at $\Delta t=0.5$, compared with $0.4611$ and $0.4560$ for DynGMA) and correspondingly yields the lowest prediction NRMSE at $\Delta t=0.5$ ($0.3540$) and matches NeuralODE at $\Delta t=0.25$ ($0.2612$ versus $0.2607$). However, because each observed endpoint includes random realization noise around the conditional mean, NRMSE reaches an irreducible floor even when coefficients improve. With 1,050 pairs at $\Delta t=0.25$, TC-AdaNSDE and DynGMA produce nearly identical prediction NRMSE ($0.2556$ and $0.2531$) despite substantially different drift and covariance errors ($\operatorname{rRMSE}_f = 0.2761$ versus $0.1707$; $\operatorname{rRMSE}_Q = 0.4022$ versus $0.2721$). Furthermore, deterministic NeuralODE achieves competitive NRMSE across all settings despite having zero diffusion ($\operatorname{rRMSE}_Q = 1$). Evaluating mean predictions alone would therefore fail to detect missing or poorly calibrated diffusion.}

\revnew{Second, transition MMD$^2$ and autonomous state MMD$^2$ expose the stochastic fidelity over short and long times that NRMSE misses. Because NeuralODE lacks diffusion, its autonomous trajectories collapse onto deterministic paths, producing state MMD$^2$ errors an order of magnitude worse than the stochastic models across every budget and interval (Figure~\ref{fig:known_nonlinear_data_sizes}e,f). Conversely, when a stochastic model overestimates covariance with sparse data (as in DynGMA at 100 pairs, with $\operatorname{rRMSE}_Q = 2.5179$ at $\Delta t=0.25$ and $1.9555$ at $\Delta t=0.5$), transition MMD$^2$ immediately penalizes the distorted distribution over one interval ($2.71\times10^{-3}$ for DynGMA versus $6.81\times10^{-4}$ for TC-AdaNSDE at $\Delta t=0.25$, where TC-AdaNSDE has $\operatorname{rRMSE}_Q = 0.6524$). Transition MMD$^2$ and state MMD$^2$ also capture distinct timescales: at 100 pairs, TC-AdaNSDE achieves the lowest transition MMD$^2$, whereas the positive neural SDE achieves slightly lower state MMD$^2$ over $\Delta t =0.5$. As data expand to 300 and 1,050 pairs, DynGMA's transition approximation with multiple steps recovers both drift and diffusion more closely and leads both MMD$^2$ metrics.}

\revnew{Together, these controlled tests justify our evaluation protocol for systems with unknown governing equations. No single observable metric is sufficient: conditional mean NRMSE tests local systematic drift, transition MMD$^2$ tests whether the combined drift and diffusion reproduce the local spread of changes over one interval, and state MMD$^2$ tests whether autonomous rollouts remain stable and populate the correct regions of state space. At the same time, the results with 100 training pairs confirm that our local moment and topology guidance prevents the severe diffusion misestimation seen in unguided likelihood fitting when transitions are scarce.}

\section{Turbulence and zonal flow dynamics}\label{sec:turbulence_zonal_flow_dynamics}

\subsection{Energy observables and trajectory sampling}\label{sec:hw_energy_sampling}

\begin{figure*}[!htbp]
\centering
\includegraphics[width=\linewidth,height=0.68\textheight,keepaspectratio]{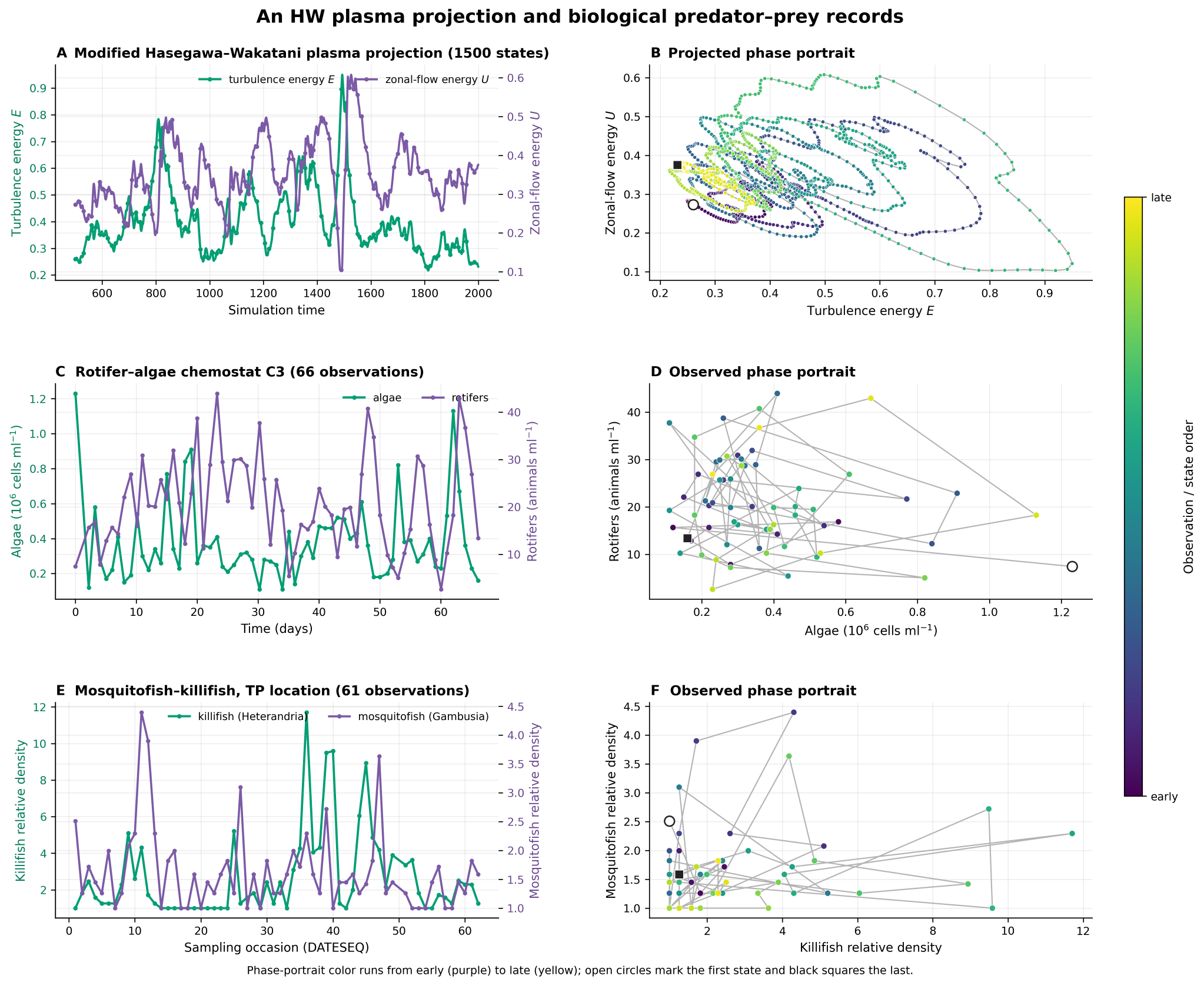}
{\ifshownewrevision\color{black}\fi
\caption{\label{fig:observed_population_hw_records}A simulated HW energy projection and observed population records. Panels A and B show 1,500 retained turbulence and zonal flow energy states from one test HW simulation. Panels C and D show 66 paired rotifer and algae observations from chemostat C3 over the first 67 days \cite{blasius2020}. Panels E and F show 61 paired Least Killifish and Eastern Mosquitofish observations at TP \cite{gatto2024}. Exponentiated log densities are shown on a positive scale of relative density, and the horizontal axis in E indexes sampling occasions. Left panels show the two variables over time or sampling order, with separate vertical axes. Right panels show their joint state trajectories. Color progresses from early (purple) to late (yellow), with an open circle marking the first state and a black square the last. The panels show the observations and simulation data used to learn the stochastic interactions.}}
\end{figure*}

\revnew{Having established that conditional-mean NRMSE, transition MMD$^2$, and state MMD$^2$ jointly diagnose drift accuracy and stochastic fidelity, we now test whether TC-AdaNSDE provides a useful effective model for a high-dimensional physical system when observations are scarce. In modified Hasegawa--Wakatani (HW) plasma turbulence, spatial density and electrostatic potential fields evolve nonlinearly across many spatial modes (Appendix~\ref{sec:hw_simulation_observables}). Projecting these fields onto two macroscopic observables---the turbulence energy $E$ and the zonal flow kinetic energy $U$, shown alongside the biological records in Figure~\ref{fig:observed_population_hw_records} (Panels A and B)---discards the spatial mode phases and correlations responsible for nonlinear energy transfer. Because distinct spatial field configurations can share the same $(E,U)$ pair yet undergo different subsequent energy changes, no exact closed two-variable equation exists. A useful reduced model must therefore capture both the average predator--prey energy exchange and the state-dependent fluctuations induced by unresolved spatial dynamics.}

\revnew{Although the true two-variable coefficients are unknown, independent full-field simulations provide ground-truth test trajectories, allowing us to apply the complete evaluation suite from Section~\ref{sec:known_equation_controls}---including four-interval drift-only forecasts and long autonomous rollouts for state MMD$^2$. We perform five independent fits using disjoint groups of five training trajectories drawn from 50 full-field simulations, evaluated against five common validation trajectories and 20 common test trajectories (49,980 test transitions at stride four, $\Delta t=1$ in normalized plasma time). Within each training group, nested maximin selections provide budgets of 25, 50, 100, 200, and 400 retained fitting transitions spanning the observed energy range, supplemented by 200 validation transitions (Appendix~\ref{sec:hw_training_groups_metrics}).}

\subsection{\revnew{Utility of TC-AdaNSDE in the sparse-data plasma regime}}\label{sec:hw_prediction_energy_distributions}

\begin{figure*}[!htbp]

\centering

\includegraphics[width=\linewidth,height=0.68\textheight,keepaspectratio]{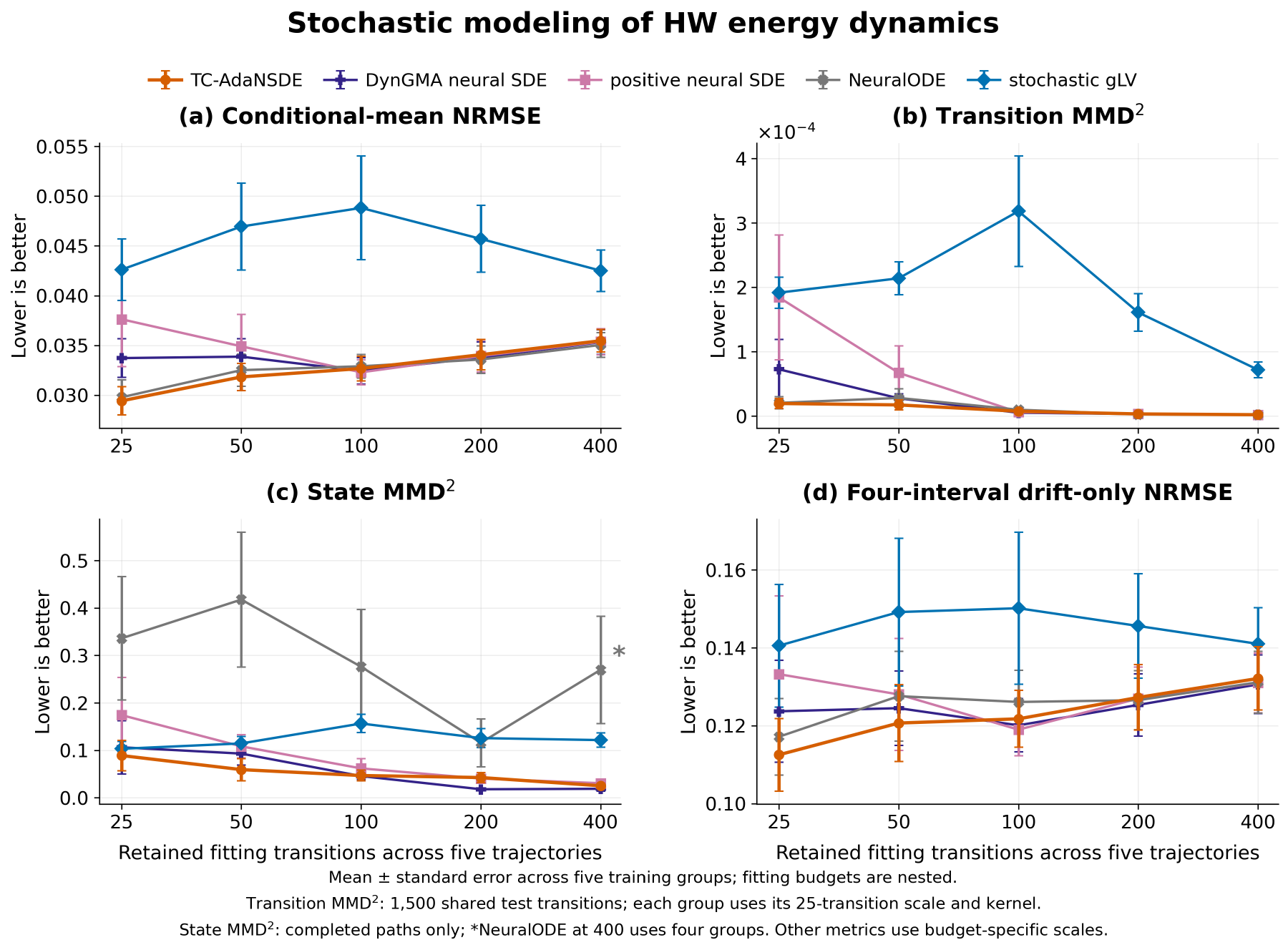}

\caption{\label{fig:hw_prediction_distribution_performance}Prediction and stochastic simulation of turbulence and zonal flow energies from HW simulations. The four panels assess (a) average changes over one interval, (b) the joint distribution of observed and simulated start--end pairs, (c) the distribution of autonomously visited states, and (d) drift only forecasts over four intervals. The five models are TC-AdaNSDE, DynGMA, positive neural SDE, stochastic gLV, and NeuralODE. Symbols show means and bars one standard error across five disjoint training groups evaluated on common test data; the starred NeuralODE state MMD$^2$ at 400 transitions uses four groups because one has no completed path. Curves join nested fitting budgets. Conditional-mean NRMSE uses all 49,980 test transitions; transition MMD$^2$ uses the same balanced subset of 1,500 transitions for every fit. Both use 64 realizations and four integration substeps per interval. Transition MMD$^2$ uses a scale and kernel fixed from the 25-transition training set in each group; normalization of the other metrics varies across budgets. State MMD$^2$ uses completed autonomous rollouts. Appendix~\ref{sec:hw_energy_model_details} gives the evaluation details; Appendix~\ref{sec:hw_guidance_ablation} compares model structures and then tests the contributions of the moment and topology losses.}

\end{figure*}

\revnew{When very few transitions are available to characterize the turbulence--zonal flow cycle (25 and 50 retained transitions), TC-AdaNSDE outperforms all four comparison models across nearly every evaluation metric simultaneously (Figure~\ref{fig:hw_prediction_distribution_performance}). At the 25-transition budget, TC-AdaNSDE achieves the lowest conditional-mean NRMSE ($0.0294$), the lowest four-step drift-only forecast error ($0.1125$), the lowest transition MMD$^2$ ($1.93\times10^{-5}$), and the lowest autonomous state MMD$^2$ ($0.0890$). By comparison, DynGMA yields $0.0337$, $0.1237$, $7.24\times10^{-5}$, and $0.1064$ on these four metrics, respectively. The same uniform advantage holds at 50 transitions, where TC-AdaNSDE again leads conditional-mean NRMSE ($0.0318$ versus $0.0339$ for DynGMA), four-step drift error ($0.1207$ versus $0.1245$), transition MMD$^2$ ($1.71\times10^{-5}$ versus $2.73\times10^{-5}$), and state MMD$^2$ ($0.0594$ versus $0.0931$).}

\revnew{These results highlight three practical advantages of TC-AdaNSDE for reduced physical modeling from limited data:}
{\ifshownewrevision\color{black}\fi
\begin{enumerate}
\item \textit{Capturing nonlinear interactions beyond classical parametric laws:} Stochastic gLV exhibits higher conditional-mean NRMSE, four-step drift error, and state MMD$^2$ than TC-AdaNSDE at every fitting budget, confirming that a rigid affine per-capita interaction law cannot adequately represent the macroscopic plasma energy transfer.
\item \textit{Reproducing macroscopic energy distributions where deterministic reduction fails:} As anticipated by the metric analysis in Section~\ref{sec:known_equation_controls}, short-horizon drift accuracy does not guarantee realistic long-term behavior. At 25 transitions, NeuralODE achieves one-step NRMSE ($0.0298$) and four-step drift error ($0.1172$) close to TC-AdaNSDE, yet its autonomous trajectories fail to reproduce the macroscopic energy distribution, yielding a state MMD$^2$ nearly four times larger ($0.3360$ versus $0.0890$; Table~\ref{tab:D2}). TC-AdaNSDE captures both the short-term average energy exchange and the long-term occupancy of energy states.
\item \textit{Preventing numerical blow-up during autonomous simulation:} Without saturation bounds and local guidance, flexible neural SDEs trained on sparse transitions frequently diverge during long autonomous rollouts. At 25 and 50 transitions, the unguided positive neural SDE completes only 39/60 and 41/60 attempted 768-step rollouts (Table~\ref{tab:D1}) and suffers large state MMD$^2$ ($0.1742$ at 25 transitions) even among its surviving paths. In contrast, TC-AdaNSDE completes all 60/60 autonomous rollouts at both 25 and 50 transitions while maintaining the closest distributional agreement with the full spatial simulations.
\end{enumerate}}

\revnew{As the fitting budget expands to 100--400 transitions, data coverage becomes dense enough for unguided neural models to constrain their coefficients directly: DynGMA achieves the lowest state MMD$^2$ from 100 transitions onward, while NeuralODE and the positive neural SDE reach slightly lower one-step NRMSE. Structural comparisons and matched loss ablations in Appendix~\ref{sec:hw_guidance_ablation} confirm that TC-AdaNSDE's advantage in the sparse-data regime stems directly from combining bounded state-scaled architecture with adaptive local moment guidance (which halves state MMD$^2$ at 25 transitions) and topology guidance (which further reduces transition and state MMD$^2$ at 50 transitions). Thus, TC-AdaNSDE is specifically effective when constructing a stable, distributionally faithful reduced model from sparse physical observations.}

\section{Biological applications}\label{sec:biological_applications}

\subsection{Population records and fits with limited observations}\label{sec:biological_records_training}

\revnew{We next test whether TC-AdaNSDE is useful for empirical ecological records, where observations come directly from laboratory experiments and field surveys rather than numerical simulations. These datasets introduce practical challenges absent in the plasma simulations: the records are short, observation intervals vary or include missing months, measurement noise is present, and no independent test trajectories exist to define a long-term stationary distribution. We evaluate two contrasting systems shown in Figure~\ref{fig:observed_population_hw_records} (Panels C--F): laboratory rotifer and algae oscillations in chemostat C3 \cite{blasius2020}, and monthly field densities of Least Killifish and Eastern Mosquitofish across three wetland sites (TP, WRGP, and WRHW) \cite{gatto2024}.}

\revnew{Because these short records do not establish a long-term stationary state distribution, we evaluate the models using conditional-mean NRMSE and transition MMD$^2$ over observed intervals. As justified in Section~\ref{sec:known_equation_controls}, examining these two metrics together tests whether a model captures both systematic population changes and the state-dependent variability of transitions without requiring long autonomous rollouts.}

\revnew{To preserve temporal ordering while testing data scarcity, each record is partitioned into contiguous temporal blocks assigned to training, validation, and testing with comparable state-space coverage (Appendix~\ref{sec:biological_observations_temporal_groups}). For the rotifer--algae chemostat, the first 67 days are divided into ten 6.7-day blocks (roughly one oscillation cycle each); the short training setting uses just two cycle blocks (12 transitions), while a longer nested setting uses five blocks (28 transitions), evaluated against 11 validation and 17 test transitions. For the fish surveys, data are grouped into six-month blocks at each site; the short training setting uses two blocks per site (equivalent to one year of monthly observations, or 30 transitions across the three sites), and the longer setting uses four blocks per site (59 transitions), evaluated against 29 validation and 54 test transitions. Fish models are fitted separately to each site before pooling standardized prediction residuals and averaging site transition MMD$^2$ scores (Appendix~\ref{sec:biological_transition_metric_longer_records}).}

\subsection{\revnew{Utility of TC-AdaNSDE on short biological records}}\label{sec:biological_prediction_transition_statistics}

\revnew{Across both ecological systems, the short training regime (12 chemostat transitions or one year of monthly fish surveys per site) demonstrates why a practical ecological model must balance drift accuracy with transition variability---and shows that TC-AdaNSDE is the only evaluated method that consistently achieves both (Figure~\ref{fig:biological_prediction_performance}; numerical values in Appendix~\ref{sec:biological_transition_metric_longer_records}, Table~\ref{tab:2}).}

\revnew{In the two-cycle rotifer--algae test, deterministic and unguided stochastic baselines fail on complementary aspects of the dynamics. Deterministic NeuralODE fails to capture the sharp cyclic turns from 12 transitions, yielding a median NRMSE of $2.656$, while DynGMA yields both higher prediction error ($1.019$) and poor transition MMD$^2$ ($0.0656$), reflecting the difficulty of fitting a multi-step Gaussian mixture likelihood to only 12 pairs. Parametric stochastic gLV is also less accurate in both NRMSE ($0.968$) and transition MMD$^2$ ($0.0313$). While the unguided positive neural SDE achieves a low median NRMSE ($0.894$), TC-AdaNSDE matches this predictive accuracy ($0.896$) while reducing transition MMD$^2$ by 26\% ($0.0188$ versus $0.0255$, with non-overlapping interquartile ranges $[0.0186, 0.0190]$ versus $[0.0248, 0.0266]$; Table~\ref{tab:2}).}

\revnew{In the one-year fish survey test, the advantage shifts to systematic prediction across sites. Here, TC-AdaNSDE achieves the lowest pooled conditional-mean NRMSE among all five models ($0.966$, interquartile range $[0.963, 0.976]$), outperforming the positive neural SDE ($0.989$), NeuralODE ($1.036$), stochastic gLV ($1.048$), and DynGMA ($1.140$). At the same time, TC-AdaNSDE maintains the second-lowest transition MMD$^2$ ($0.0265$), close to DynGMA ($0.0239$) and ahead of the positive neural SDE ($0.0281$), stochastic gLV ($0.0279$), and NeuralODE ($0.0335$). Thus, whereas DynGMA sacrifices mean prediction accuracy on both short biological datasets and the positive neural SDE loses distributional precision on the chemostat cycles, TC-AdaNSDE delivers reliable mean predictions and well-calibrated transition distributions across both laboratory and field regimes.}

\begin{figure*}[!htbp]

\centering

\includegraphics[width=\linewidth,height=0.68\textheight,keepaspectratio]{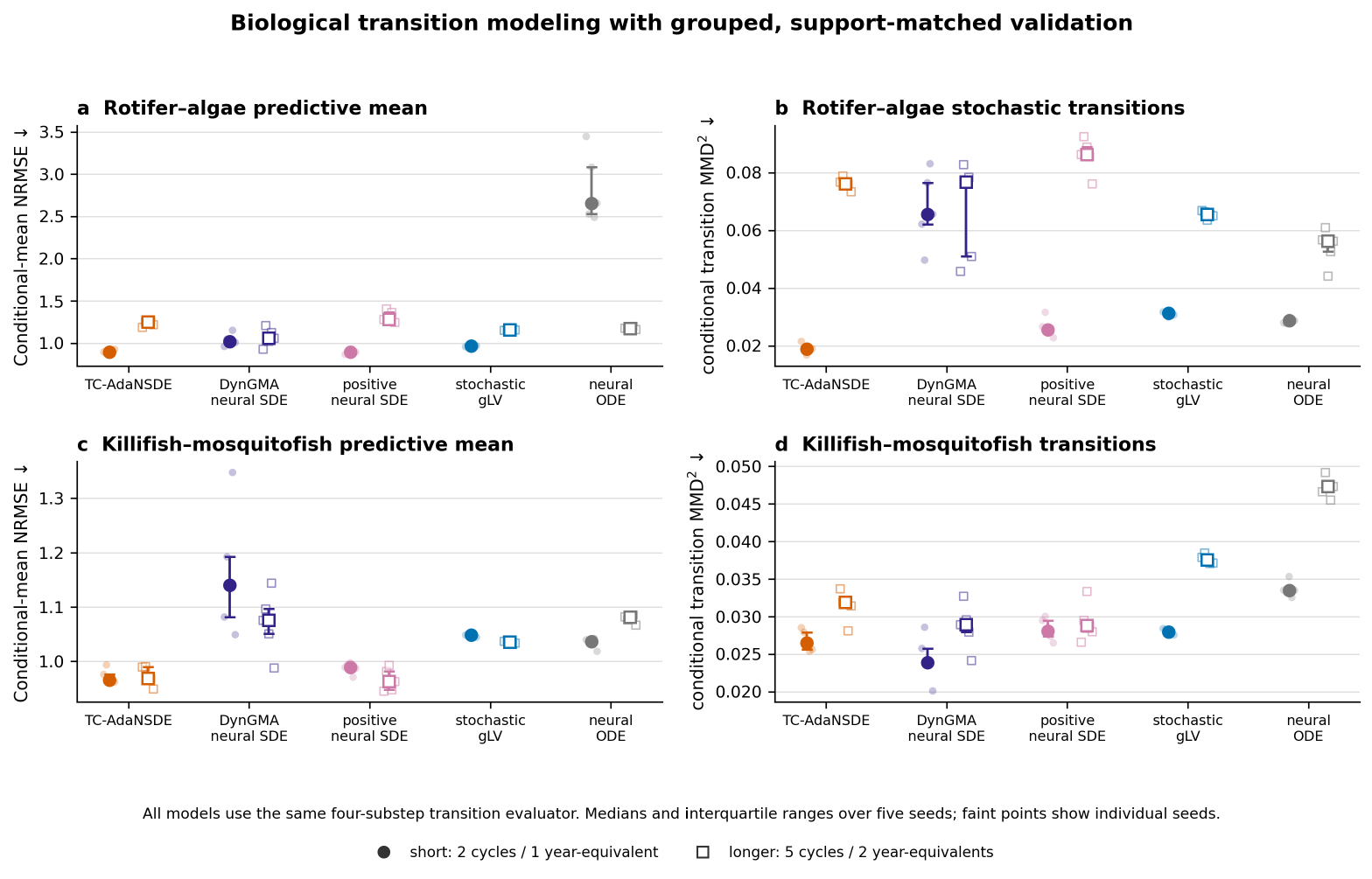}

\caption{\label{fig:biological_prediction_performance}Predictive and distributional agreement as biological training information changes. Conditional mean NRMSE and transition MMD$^2$ assess complementary aspects of the fitted dynamics. DynGMA adds a reference based on a Gaussian mixture transition likelihood. All five models are evaluated with the same solver with four substeps, using 64 endpoint sets to estimate conditional means and distributional scores. Filled circles and open squares indicate the short and longer training settings. Large symbols show medians, bars interquartile ranges, and faint points individual optimizer seeds. Tables~\ref{tab:2} and \ref{tab:B2} in Appendix~\ref{sec:biological_transition_metric_longer_records} give the numerical results.}

\end{figure*}

\revnew{When the training records are extended to five chemostat cycles (28 transitions) and two years of fish surveys (59 transitions), unguided neural models gain enough data to match or surpass the guided fit: DynGMA and NeuralODE lead the longer rotifer metrics, while the positive neural SDE leads the two-year fish metrics (Figure~\ref{fig:biological_prediction_performance} and Table~\ref{tab:B2}). Because adding temporal blocks in real ecological records alters the state-space distribution of training transitions as well as their sample count, errors do not decrease monotonically with record length. This pattern reinforces the conclusion from the plasma experiments: TC-AdaNSDE provides its greatest practical value precisely when empirical records are shortest (one to two oscillation cycles or a single annual cycle).}

\subsection{\revnew{Interpretable state-dependent interaction and fluctuation fields}}\label{sec:biological_learned_drift_fluctuations}

\begin{figure*}[!htbp]

\centering

\includegraphics[width=\linewidth,height=0.68\textheight,keepaspectratio]{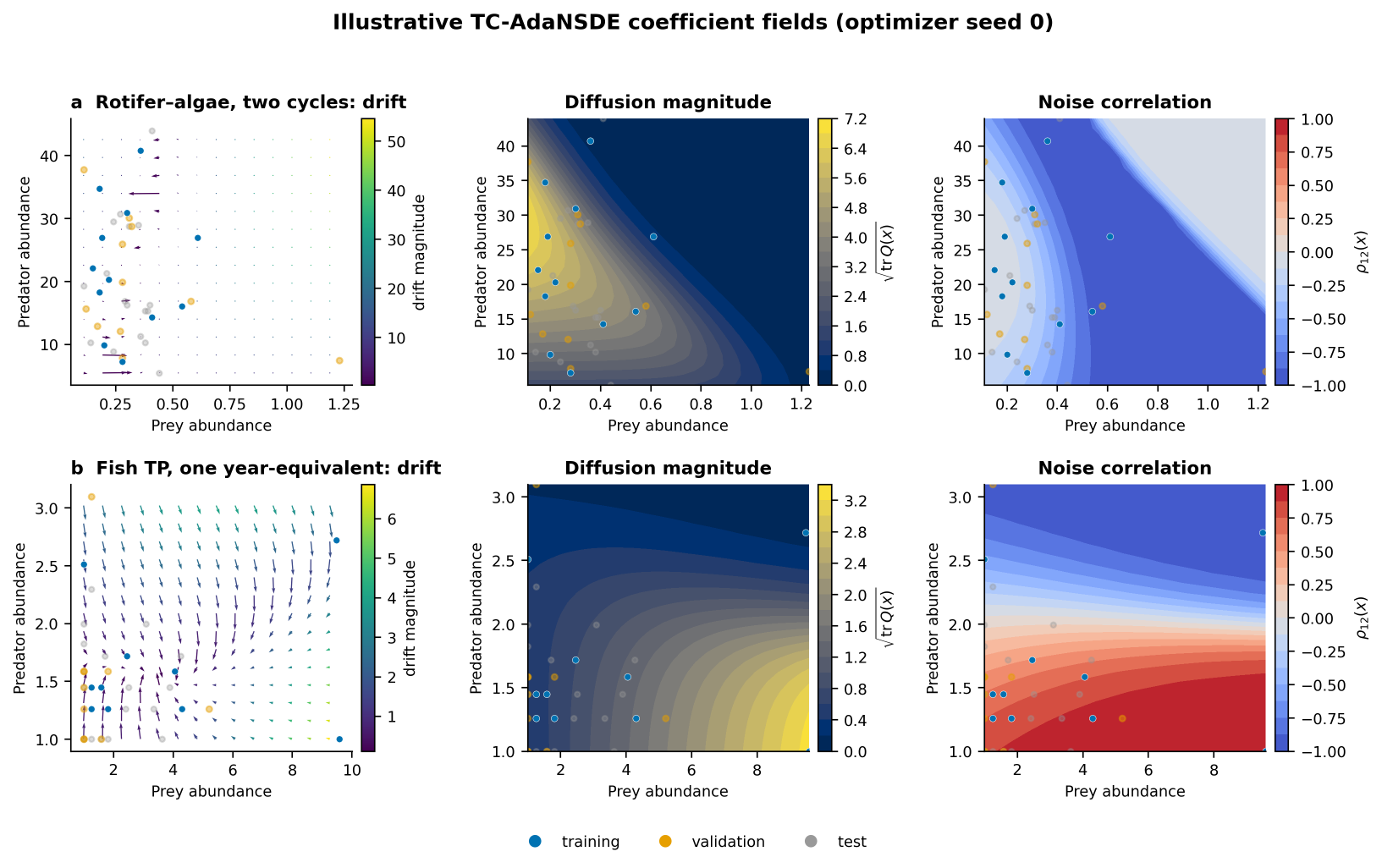}

\caption{\label{fig:biological_learned_coefficient_fields}State dependence of the learned biological interaction and fluctuations. The fields illustrate what the neural coefficients can represent beyond a fixed interaction law. Rows show fits with seed zero to C3 and the TP fish site. Columns show drift, total fluctuation amplitude $\sqrt{\operatorname{tr}Q}$, and noise correlation $\rho=Q_{12}/\sqrt{Q_{11}Q_{22}}$. Arrow direction is normalized and color denotes drift magnitude. Training, validation, and test starts are overlaid. The grid spans the observed coordinate ranges, including sparsely sampled regions where field values rely on neural interpolation or extrapolation.}

\end{figure*}

\revnew{Beyond quantitative prediction errors, TC-AdaNSDE turns sparse transition pairs into a continuous, physically interpretable phase-space portrait of the ecological interaction. Figure~\ref{fig:biological_learned_coefficient_fields} displays the learned drift vector field $f_\theta(x)$, total fluctuation amplitude $\sqrt{\operatorname{tr}Q_\theta(x)}$, and instantaneous noise correlation $\rho_\theta(x)$ fitted to the short records for chemostat C3 and the TP fish site.}

\revnew{Even from 12 transitions (C3) or 10 site transitions (TP), the guided networks recover smooth, biologically plausible phase-space geometry without wild extrapolation in unsampled corners. The drift arrows trace  predominantly counterclockwise predator--prey circulation around the coexistence region while respecting positivity near the boundaries. Moreover, the learned diffusion fields reveal system-specific stochastic structure that constant-correlation or deterministic models cannot represent: in the chemostat (Figure~\ref{fig:biological_learned_coefficient_fields}a), fluctuations are predominantly negatively correlated ($\rho_{12} < 0$), whereas at the TP fish site (Figure~\ref{fig:biological_learned_coefficient_fields}b), the inferred correlation transitions smoothly from negative at higher predator densities to positive at lower predator densities. TC-AdaNSDE thus provides not only accurate short-record transitions, but also a coherent continuous map of systematic tendencies and coupled environmental fluctuations across state space.}

\section{Discussion and limitations}\label{sec:discussion_limitations}

\revnew{Constructing an effective stochastic model from limited observations requires both flexible coefficient representations and a reliable way to evaluate them when true generating equations are unknown. Our progression across three experimental settings addresses both requirements. First, the known-equation nonlinear control in Section~\ref{sec:known_equation_controls} justifies our multi-metric evaluation framework: because models with poor diffusion recovery (or zero diffusion, as in NeuralODE) can still achieve low one-step conditional-mean NRMSE, NRMSE must be paired with transition MMD$^2$ and autonomous state MMD$^2$ to verify that a learned model captures both systematic drift and stochastic variability. Second, the HW plasma turbulence reduction (Section~\ref{sec:turbulence_zonal_flow_dynamics}) and empirical biological records (Section~\ref{sec:biological_applications}) demonstrate the practical utility of TC-AdaNSDE under severe data scarcity. Across both physical and biological domains, when observations are restricted to 25--50 plasma transitions, two chemostat cycles, or one year of monthly field surveys, TC-AdaNSDE outperforms parametric gLV, deterministic NeuralODE, and unguided neural SDE baselines by combining accurate short-horizon predictions with well-calibrated transition and state distributions---and by preventing the rollout divergence that afflicts unguided neural SDEs.}

\revnew{The method achieves this sample efficiency by combining bounded state-scaled neural coefficients with two forms of guidance: adaptive local moment estimates pool nearby transitions to separate local drift from covariance, while soft extinction and coexistence constraints regularize the drift where data are sparse. As the available training data expand, the relative advantage of these structural priors diminishes and unguided methods---particularly DynGMA with its multi-step Gaussian mixture likelihood, or the positive neural SDE---match or overtake TC-AdaNSDE on several metrics. The HW ablations in Appendix~\ref{sec:hw_guidance_ablation} clarify how the components interact in the sparse regime: adaptive moment guidance drives the primary reduction in autonomous state-distribution error at both 25 and 50 transitions, while adding topology guidance provides further distributional gains at 50 transitions.}

\revnew{Several limitations define the scope of our findings. First, the inferred drift and diffusion represent effective dynamics at the observation interval $\Delta t$; finite sampling intervals, measurement noise, and unresolved degrees of freedom (such as spatial plasma modes or unmeasured environmental covariates) are absorbed into these effective coefficients. In particular, pooling biological increments with unequal durations via a median reference interval $\Delta t_{\mathrm{ref}}$ makes the local moment targets approximate. Second, our evaluation budgets reflect retained transitions under specific sampling designs: maximin state-space selection from multiple trajectories in HW (plus validation data), and support-matched temporal block assignments in the retrospective biological tests. Third, the soft phase-space penalties guide local deterministic derivatives near extinction and a chosen coexistence anchor without guaranteeing global phase-space topology or long-time stochastic stability outside the sampled domain. Extending this framework to systems with strong non-Markovian memory or multiple competing attractors will require adapting both the effective state variables and the dynamical priors.}

\section{Conclusion}\label{sec:conclusion}

\revnew{We have developed TC-AdaNSDE, a framework for learning effective two-variable predator--prey stochastic differential equations from sparse observations by integrating bounded state-scaled neural networks, correlated diffusion, adaptive local moment guidance, and soft extinction and coexistence constraints. Using a nonlinear stochastic system with known coefficients, we first established that evaluating learned SDEs requires combining conditional-mean NRMSE with transition MMD$^2$ and autonomous state MMD$^2$, because single-step mean prediction error alone cannot detect missing or miscalibrated diffusion. Applying this metric suite to reduced turbulence and zonal flow energies from Hasegawa--Wakatani plasma simulations and to short experimental and field population records, we demonstrated that TC-AdaNSDE is consistently useful in the small-sample regime: it achieves the lowest errors across all four predictive and distributional metrics at 25 and 50 plasma transitions with 100\% autonomous rollout completion, and provides the most reliable balance of prediction accuracy and transition distribution fidelity across two-cycle plankton chemostat records and one-year fish surveys. Together, these results show that pairing local statistical pooling with qualitative dynamical structure enables flexible neural SDEs to extract useful, interpretable stochastic dynamics from severely limited physical and biological data.}

% Before submission, supply the applicable author-approved statements here.
% \begin{acknowledgments}
% Funding and acknowledgments.
% \end{acknowledgments}
% \section*{Author declarations}
% \subsection*{Conflict of interest}
% Insert the authors' conflict-of-interest disclosure.
% \subsection*{Author contributions}
% Insert the agreed CRediT contribution statement for every author.
% \section*{Data availability}
% Insert the data availability statement and repository links, as applicable.

\begin{acknowledgments}
This work is supported in part by the Ministry of
Education (MOE) AcRF Tier 1 grants RS02/23 and
RG156/23, and the National Research Foundation
Singapore (NRF) core funding ``Fusion Science for
Clean Energy'', under the framework of the Singapore
Alliance with France for Fusion Energy (SAFE).
The computational work for this article was partially
performed on resources of the National Supercomputing
Centre (NSCC), Singapore (Project 12003525).
\end{acknowledgments}

\section*{Author Declarations}

\subsection*{Conflict of Interest}

The authors have no conflicts to disclose.

% \subsection*{Author Contributions}

% [Authors to supply author contributions.]

\section*{Data Availability}

The code and data that support the findings of this study are available from the corresponding author upon reasonable request.

\appendix

\section{Training objective and optimization}\label{sec:training_optimization_details}

Sections~\ref{sec:adaptive_local_moment_guidance} and \ref{sec:joint_objective_regularization} define the regression objective and loss normalizations. For a fitted neighborhood radius $h(q)$, the weights are proportional to $\exp[-\|(x_n-q)/s_x\|_2^2/(2h(q)^2)]$, where $s_x$ is the vector of training-start standard deviations with a small positive offset. The covariance target $\widehat Q(q)$ is symmetrized with eigenvalues floored at $10^{-8}$. Biological moment estimates use density-adapted neighborhoods with candidate neighbor counts 8, 12, and 20; HW and nonlinear-control estimates use nearest-neighbor local linear regression with candidate counts 10, 20, 40, and 80, selected by Gaussian validation likelihood.

The phase-space penalty weights in Section~\ref{sec:phase_space_guidance} are $\lambda_{\mathrm{s}}=0.004$, $\lambda_{\mathrm{fp}}=0.006$, and $\lambda_{\mathrm{st}}=0.008$, with margin $\delta=10^{-4}$. The coefficient regularization weights in Section~\ref{sec:joint_objective_regularization} are $\lambda_f=0.002$, $\lambda_Q=0.03$, and $\lambda_\rho=0.001$, with correlation $\rho_\theta = Q_{12}/\sqrt{\max(Q_{11}Q_{22},10^{-12})}$. During simulation, state components are bounded below by $10^{-6}$ after each update.

Separate drift and diffusion networks use two SiLU hidden layers (32 units per layer for biology and the nonlinear control; 64 units for HW) and are trained with AdamW (learning rate $3\times10^{-4}$, weight decay $5\times10^{-5}$, batch size 128, and gradient clipping at $0.25$). The phase-space schedule $\alpha(e)$ starts at epoch 200 and ramps linearly over 300 epochs for biology (checkpoint selected by validation loss among epochs 600--900), and starts at epoch 120 over 240 epochs for HW and the nonlinear control (selected among epochs 360--600, early stopping patience 160).

\revadd{We implement DynGMA using Algorithm 2 of Zhu and Li \cite{zhu2024}, with two propagation steps per recorded interval.} Five weighted sigma points in two dimensions ($\lambda=1$, weights $1/3$ at the center and $1/6$ at four displaced points) propagate the midpoint Gaussian into a five-component Gaussian mixture. DynGMA shares the positive neural SDE architecture, normalization, coefficient regularization, and optimizer settings, selecting checkpoints by its validation likelihood from epoch 1 (maximum 900 epochs with patience 300 for biology; 600 epochs with patience 160 for HW and the nonlinear control).

\section{\revadd{Nonlinear control: sampling and stochastic simulation}}\label{sec:nonlinear_control_simulation}

\revadd{For each of three seeds, 30 trajectories start independently and uniformly in $[0.2,2]^2$ and evolve via Euler--Maruyama with step $0.005$ and a componentwise floor of $10^{-6}$. After a burn-in of three time units, states are stored every $0.05$ for 120 time units (five fitting, five validation, and 20 test trajectories). Random selection among candidate starts separated by $0.5$ time units yields 1,050 fitting pairs (whose first 100 and 300 entries form the nested subsets), alongside 225 validation and 225 test pairs shared across $\Delta t=0.25$ and $0.5$.}

\revadd{All evaluation sizes use the normalization and kernel of the 100-pair subset within each seed. Transition MMD$^2$ uses all 225 test pairs and Gaussian bandwidths $0.5$, 1, and 2 times the median pairwise training distance. Conditional-mean NRMSE and transition MMD$^2$ average 64 realizations per test start integrated with step $0.01$. For state MMD$^2$, 12 autonomous paths per seed run for 60 time units with step $0.01$ (storing up to 600 states every $0.05$ after a five-unit burn-in, stopping if components become nonfinite or reach 100) and are compared against 600 reference states using Gaussian bandwidths $0.05, 0.1, 0.2, 0.5,$ and 1. Out of 36 attempted autonomous paths across the three seeds, TC-AdaNSDE completes 34 and 35 paths at 300 and 1,050 pairs ($\Delta t=0.25$), DynGMA completes 35 at 100 pairs ($\Delta t=0.5$), TC-AdaNSDE and the positive neural SDE complete 35 at 300 pairs ($\Delta t=0.5$), and all other settings complete 36/36. Table~\ref{tab:nonlinear_coefficient_errors} reports the relative coefficient recovery errors.}

\begin{table*}[!htbp]
\centering
{\ifshowrevision\color{black}\fi
\caption{Relative drift and covariance errors for the nonlinear system, using the same fits as Figure~\ref{fig:known_nonlinear_data_sizes}. Values are means $\pm$ one standard error across three repeats. Equation~\eqref{eq:relative_coefficient_rmse} is evaluated in the original state coordinates at the same 225 test starting states per repeat. Lower values indicate closer coefficient recovery.}
\label{tab:nonlinear_coefficient_errors}
\fontsize{9}{11}\selectfont
\setlength{\tabcolsep}{4pt}
\renewcommand{\arraystretch}{1.15}
\begin{tabular}{lcccccc}
\toprule
Model & \multicolumn{2}{c}{100 training pairs} & \multicolumn{2}{c}{300 training pairs} & \multicolumn{2}{c}{1,050 training pairs} \\
\cmidrule(lr){2-3}\cmidrule(lr){4-5}\cmidrule(lr){6-7}
 & $\operatorname{rRMSE}_f$ & $\operatorname{rRMSE}_Q$ & $\operatorname{rRMSE}_f$ & $\operatorname{rRMSE}_Q$ & $\operatorname{rRMSE}_f$ & $\operatorname{rRMSE}_Q$ \\
\midrule
\multicolumn{7}{l}{$\Delta t=0.25$} \\
TC-AdaNSDE & $0.3125\pm0.0821$ & $0.6524\pm0.0301$ & $0.2596\pm0.0649$ & $0.4536\pm0.0850$ & $0.2761\pm0.0846$ & $0.4022\pm0.0594$ \\
DynGMA & $0.4611\pm0.0830$ & $2.5179\pm1.2146$ & $0.1773\pm0.0209$ & $0.3551\pm0.1399$ & $0.1707\pm0.0413$ & $0.2721\pm0.0549$ \\
Positive neural SDE & $0.3286\pm0.0590$ & $0.8506\pm0.1930$ & $0.1839\pm0.0398$ & $0.4283\pm0.1088$ & $0.1955\pm0.0377$ & $0.3738\pm0.0488$ \\
Stochastic gLV & $0.4842\pm0.0247$ & $0.4751\pm0.0076$ & $0.3399\pm0.0860$ & $0.2385\pm0.0182$ & $0.3362\pm0.0845$ & $0.1728\pm0.0120$ \\
NeuralODE & $0.2957\pm0.0377$ & $1.0000$ & $0.2086\pm0.0394$ & $1.0000$ & $0.1920\pm0.0345$ & $1.0000$ \\
\midrule
\multicolumn{7}{l}{$\Delta t=0.5$} \\
TC-AdaNSDE & $0.3115\pm0.0366$ & $0.6367\pm0.0294$ & $0.2844\pm0.0437$ & $0.4929\pm0.0647$ & $0.2548\pm0.0285$ & $0.4914\pm0.0677$ \\
DynGMA & $0.4560\pm0.1138$ & $1.9555\pm1.3899$ & $0.2620\pm0.1300$ & $0.3020\pm0.0878$ & $0.1409\pm0.0617$ & $0.2943\pm0.0489$ \\
Positive neural SDE & $0.3847\pm0.0457$ & $0.6305\pm0.0730$ & $0.3061\pm0.0759$ & $0.4331\pm0.0757$ & $0.2751\pm0.0322$ & $0.4617\pm0.0731$ \\
Stochastic gLV & $0.4534\pm0.0327$ & $0.4743\pm0.0059$ & $0.3420\pm0.0760$ & $0.2661\pm0.0187$ & $0.3561\pm0.0660$ & $0.2435\pm0.0159$ \\
NeuralODE & $0.3173\pm0.0245$ & $1.0000$ & $0.2728\pm0.0426$ & $1.0000$ & $0.2459\pm0.0327$ & $1.0000$ \\
\bottomrule
\end{tabular}
}
\end{table*}

\section{From HW fields to an effective energy model}\label{sec:hw_energy_model_details}

\subsection{Simulation and observables}\label{sec:hw_simulation_observables}

Following the modified HW formulation in \cite{huang2026}, the density perturbation $n(x,y,t)$ and electrostatic potential $\phi(x,y,t)$ (with vorticity $\zeta=\nabla^2\phi$) evolve on a periodic domain $[0,L_x]\times[0,L_y]$ according to
\begin{equation}
\begin{aligned}
\partial_t\zeta+\{\phi,\zeta\}
 &=\alpha(\widetilde\phi-\widetilde n)-D\nabla^4\zeta-\nu\langle\zeta\rangle_y,\\
\partial_t n+\{\phi,n\}
 &=\alpha(\widetilde\phi-\widetilde n)-\kappa\partial_y\phi-D\nabla^4n,
\end{aligned}
\label{eq:modified_hw_equations}
\end{equation}
where $\{a,b\}=\partial_xa\,\partial_yb-\partial_ya\,\partial_xb$, $\langle\cdot\rangle_y$ is the poloidal average, and $\widetilde a=a-\langle a\rangle_y$ is the nonzonal component. Simulations use $\alpha=2$, $D=10^{-4}$, $\nu=0.01$, $\kappa=1.0403225806451613$, $L_x=12.875$, and $L_y=51.5$ \cite{huang2026}. The two retained energy observables are the nonzonal turbulence energy $E$ and zonal flow kinetic energy $U$:
\begin{equation}
\begin{aligned}
E&=\frac{1}{2L_xL_y}\int\left(|\nabla\widetilde\phi|^2+\widetilde n^2\right)\,dx\,dy,\\
U&=\frac{1}{2L_x}\int\left|\langle\partial_x\phi\rangle_y\right|^2\,dx.
\end{aligned}
\label{eq:hw_energy_observables}
\end{equation}
From 50 trajectories of 12,000 outputs each, the first 2,000 transient outputs are discarded and every fourth state is retained ($\Delta t=1$), providing 2,500 states per trajectory (49,980 test transitions across the 20 test trajectories).

\subsection{Training groups and numerical metrics}\label{sec:hw_training_groups_metrics}

Twenty-five trajectories form five disjoint training groups (five trajectories each, with nested maximin selections of 5, 10, 20, 40, and 80 transitions per trajectory), five trajectories supply 200 common validation transitions, and 20 trajectories supply the test set. Conditional-mean NRMSE uses all 49,980 test transitions (64 realizations, four Euler--Maruyama substeps per interval $\Delta t=1$). Transition MMD$^2$ uses a fixed balanced subset of 1,500 test transitions (75 per test trajectory) standardized by the 25-transition training starts, with Gaussian bandwidths $0.5d_{\mathrm{med}}, d_{\mathrm{med}},$ and $2d_{\mathrm{med}}$. Four-interval drift-only error integrates deterministic trajectories from 128 test windows over four intervals.

For state MMD$^2$, 12 paths per fit are initialized from test states and simulated autonomously for 768 intervals (discarding the first 20\%, subsampling up to 1,500 states, and comparing against test trajectories with Gaussian bandwidths $0.05, 0.1, 0.2, 0.5,$ and 1). A path is stopped if either energy exceeds the larger of five times its training maximum and twice its test 99.9th percentile. Table~\ref{tab:D1} reports the completed rollouts out of 60 attempts per budget.

\begin{table*}[!htbp]

\centering
\caption{\label{tab:D1}Completed HW rollouts among 60 attempts per model and fitting budget; NeuralODE rollouts are deterministic. Attempts comprise 12 paths for each of five training groups. State MMD$^2$ in Figure~\ref{fig:hw_prediction_distribution_performance} uses the completed subset in each group.}
\fontsize{9.5}{11.5}\selectfont
\setlength{\tabcolsep}{4pt}
\renewcommand{\arraystretch}{1.18}
\begin{tabularx}{\linewidth}{@{}>{\hsize=2.1\hsize\raggedright\arraybackslash}X>{\hsize=0.78\hsize\raggedright\arraybackslash}X>{\hsize=0.78\hsize\raggedright\arraybackslash}X>{\hsize=0.78\hsize\raggedright\arraybackslash}X>{\hsize=0.78\hsize\raggedright\arraybackslash}X>{\hsize=0.78\hsize\raggedright\arraybackslash}X@{}}
\toprule
\textbf{Model} & \textbf{25} & \textbf{50} & \textbf{100} & \textbf{200} & \textbf{400} \\
\midrule
TC-AdaNSDE & 60/60 & 60/60 & 60/60 & 53/60 & 60/60 \\
DynGMA neural SDE & 60/60 & 60/60 & 60/60 & 60/60 & 60/60 \\
Positive neural SDE & 39/60 & 41/60 & 60/60 & 60/60 & 60/60 \\
Stochastic gLV & 60/60 & 60/60 & 57/60 & 54/60 & 60/60 \\
NeuralODE & 60/60 & 59/60 & 60/60 & 60/60 & 48/60 \\
\bottomrule
\end{tabularx}
\end{table*}

\section{Model structure and guidance losses for HW energy dynamics}\label{sec:hw_guidance_ablation}

We use the HW turbulence and zonal flow energies to examine the contributions of model architecture and training guidance in two stages: first comparing different stochastic model structures at 25 transitions, and second holding the TC-AdaNSDE architecture fixed to ablate the local moment and topology losses at 25 and 50 transitions.

\subsection{Comparison of stochastic model structures}\label{sec:hw_structure_comparison}

Table~\ref{tab:nsde_components} summarizes the structural and guidance components of the neural SDE variants. We add an unscaled diagonal neural SDE ($Q_{12}=0$) and an unscaled full-covariance neural SDE ($\rho\neq 0$) to the five principal models at 25 transitions. Both unscaled references use two hidden layers of 64 units and Gaussian transition likelihood fitting without saturation bounds or guidance losses.

\begin{table*}[htbp]
\caption{Comparison of structural and training components across the neural SDE models. All four models use the Gaussian transition negative log likelihood ($\mathcal{L}_{\mathrm{NLL}}$).}
\label{tab:nsde_components}
\begin{ruledtabular}
\begin{tabular}{lcccc}
Model & Positive-state & Correlated & Local moment & Topology \\
 & scaling ($\tilde{x}_i$) & diffusion ($\rho \neq 0$) & guidance ($\mathcal{L}_{\mathrm{moment}}$) & guidance ($\mathcal{L}_{\mathrm{topology}}$) \\
\hline
Diagonal neural SDE        & \xmark & \xmark & \xmark & \xmark \\
Full-covariance neural SDE & \xmark & \cmark & \xmark & \xmark \\
Positive neural SDE        & \cmark & \cmark & \xmark & \xmark \\
TC-AdaNSDE                 & \cmark\footnotemark[1] & \cmark & \cmark & \cmark \\
\end{tabular}
\end{ruledtabular}
\footnotetext[1]{Includes explicit $\tanh$/$\mathrm{sigmoid}$ per-capita saturation bounds ($B_f = 2$, $B_\sigma = 0.35$).}
\end{table*}

Table~\ref{tab:D2} shows that reproducing autonomous stochastic behavior requires more than short-interval drift accuracy. TC-AdaNSDE, the diagonal reference, and the full-covariance reference have nearly identical four-step drift errors ($0.11254$, $0.11253$, and $0.11405$), yet their autonomous state MMD$^2$ scores differ sharply ($0.0890$, $0.2326$, and $0.4257$). Allowing unscaled noise correlation alone degrades both transition MMD$^2$ ($3.99\times10^{-3}$) and state MMD$^2$ ($0.4257$). TC-AdaNSDE combines the lowest local prediction and transition errors with the closest autonomous energy distribution and 60/60 rollout completion.

\begin{table*}[!htbp]

\centering
\caption{\label{tab:D2}Expanded comparison of stochastic model structures at the smallest HW fitting budget (25 transitions). Direct scores are mean $\pm$ standard error across five training groups on the common test set. Conditional-mean NRMSE uses 64 realizations and four substeps over one observation interval. Transition MMD$^2$ uses the common 1,500 test transitions and training-based kernel in Appendix~\ref{sec:hw_training_groups_metrics}; values in that column are multiplied by $10^5$. Four step error is drift only. State MMD$^2$ is conditional on completion. The last column gives completed paths out of 60.}
\fontsize{9.5}{11.5}\selectfont
\setlength{\tabcolsep}{4pt}
\renewcommand{\arraystretch}{1.18}
\begin{tabularx}{\linewidth}{@{}>{\hsize=1.15\hsize\raggedright\arraybackslash}X>{\hsize=1.05\hsize\raggedright\arraybackslash}X>{\hsize=1.05\hsize\raggedright\arraybackslash}X>{\hsize=1.05\hsize\raggedright\arraybackslash}X>{\hsize=1.05\hsize\raggedright\arraybackslash}X>{\hsize=0.65\hsize\raggedright\arraybackslash}X@{}}
\toprule
\textbf{Model} & \textbf{Conditional mean NRMSE} & \textbf{Transition MMD$^2$ $\times 10^5$} & \textbf{State MMD$^2$} & \textbf{Four step error} & \textbf{Completed} \\
\midrule
TC-AdaNSDE & 0.02942 $\pm$ 0.00143 & 1.927 $\pm$ 0.781 & 0.08902 $\pm$ 0.03196 & 0.11254 $\pm$ 0.00932 & 60/60 \\
DynGMA neural SDE & 0.03372 $\pm$ 0.00195 & 7.240 $\pm$ 4.616 & 0.10638 $\pm$ 0.05594 & 0.12370 $\pm$ 0.01305 & 60/60 \\
Positive neural SDE & 0.03763 $\pm$ 0.00475 & 18.419 $\pm$ 9.681 & 0.17423 $\pm$ 0.07922 & 0.13322 $\pm$ 0.02011 & 39/60 \\
Stochastic gLV & 0.04261 $\pm$ 0.00308 & 19.149 $\pm$ 2.414 & 0.10325 $\pm$ 0.01522 & 0.14054 $\pm$ 0.01574 & 60/60 \\
NeuralODE & 0.02978 $\pm$ 0.00177 & 2.024 $\pm$ 0.938 & 0.33602 $\pm$ 0.13005 & 0.11717 $\pm$ 0.00984 & 60/60 \\
Diagonal neural SDE & 0.02968 $\pm$ 0.00137 & 2.091 $\pm$ 0.862 & 0.23262 $\pm$ 0.12106 & 0.11253 $\pm$ 0.00719 & 60/60 \\
Full covariance neural SDE & 0.05318 $\pm$ 0.00538 & 398.655 $\pm$ 226.594 & 0.42570 $\pm$ 0.10484 & 0.11405 $\pm$ 0.00750 & 50/60 \\
\bottomrule
\end{tabularx}
\end{table*}

\subsection{Contributions of moment and topology losses}\label{sec:hw_loss_ablation}

To isolate the effects of the guidance losses, we hold TC-AdaNSDE's bounded state-scaled architecture, correlated diffusion, regularization, initialization, and minibatches fixed, and refit four variants at 25 and 50 transitions by switching the moment and topology terms on or off ($a,b\in\{0,1\}$):
\begin{equation}
\begin{aligned}
\mathcal L_{a,b}={}&0.5\mathcal L_{\mathrm{NLL}}
+a\bigl(\mathcal L_{\mathrm{moment},f}+\mathcal L_{\mathrm{moment},Q}\bigr)\\
&+b\mathcal L_{\mathrm{topology}}+\mathcal L_{\mathrm{reg}}.
\end{aligned}
\label{eq:matched_guidance_objective}
\end{equation}

Table~\ref{tab:hw_guidance_ablation} shows that adaptive local moment guidance provides the primary reduction in state-distribution and transition errors: without topology guidance, moment matching lowers mean state MMD$^2$ from $0.1350$ to $0.0714$ (and transition MMD$^2$ from $3.36\times10^{-5}$ to $1.99\times10^{-5}$) at 25 transitions, and lowers state MMD$^2$ from $0.0815$ to $0.0570$ at 50 transitions while achieving 60/60 completed rollouts. Adding topology guidance to moment matching further reduces state MMD$^2$ to $0.0444$ and transition MMD$^2$ to $1.45\times10^{-5}$ at 50 transitions, while yielding the lowest conditional-mean NRMSE and four-interval drift errors at both budgets.

The ablation fits use different random seeds from the main HW comparison in Figure~\ref{fig:hw_prediction_distribution_performance} and Table~\ref{tab:D2}.

\begin{table*}[!htbp]
\centering
\caption{\label{tab:hw_guidance_ablation}Matched HW guidance ablation. Within each fitting budget, the variants use the same bounded TC-AdaNSDE architecture and differ only in the active guidance terms. Errors are mean $\pm$ standard error across five paired training groups. Conditional-mean NRMSE averages 64 realizations integrated with four substeps over one observation interval. Transition MMD$^2$ uses the shared 1,500-pair subset and fixed training reference; values in that column are multiplied by $10^5$. Four-interval error uses drift only. State MMD$^2$ uses completed autonomous paths; completion counts are out of 60 attempts. All conditional prediction realizations completed.}
\fontsize{9.5}{11.5}\selectfont
\setlength{\tabcolsep}{3pt}
\renewcommand{\arraystretch}{1.18}
\begin{tabularx}{\linewidth}{@{}>{\hsize=0.55\hsize\raggedright\arraybackslash}X>{\hsize=0.9\hsize\raggedright\arraybackslash}X>{\hsize=1.25\hsize\raggedright\arraybackslash}X>{\hsize=1.1\hsize\raggedright\arraybackslash}X>{\hsize=1.15\hsize\raggedright\arraybackslash}X>{\hsize=1.25\hsize\raggedright\arraybackslash}X>{\hsize=0.8\hsize\raggedright\arraybackslash}X@{}}
\toprule
\textbf{Fitting transitions} & \textbf{Guidance} & \textbf{Conditional mean NRMSE} & \textbf{Transition MMD$^2$ $\times 10^5$} & \textbf{State MMD$^2$} & \textbf{Four-interval drift error} & \textbf{Completed} \\
\midrule
25 & Neither & 0.03073 $\pm$ 0.00155 & 3.358 $\pm$ 0.956 & 0.13496 $\pm$ 0.03851 & 0.11684 $\pm$ 0.00995 & 57/60 \\
25 & Moment only & 0.02953 $\pm$ 0.00148 & 1.989 $\pm$ 0.688 & 0.07141 $\pm$ 0.01712 & 0.11310 $\pm$ 0.00953 & 60/60 \\
25 & Topology only & 0.03082 $\pm$ 0.00159 & 3.487 $\pm$ 1.020 & 0.13934 $\pm$ 0.03907 & 0.11721 $\pm$ 0.01001 & 59/60 \\
25 & Both (full) & 0.02952 $\pm$ 0.00142 & 1.981 $\pm$ 0.674 & 0.09149 $\pm$ 0.03246 & 0.11270 $\pm$ 0.00918 & 60/60 \\
\midrule
50 & Neither & 0.03180 $\pm$ 0.00165 & 1.783 $\pm$ 0.895 & 0.08152 $\pm$ 0.03419 & 0.12084 $\pm$ 0.01104 & 60/60 \\
50 & Moment only & 0.03194 $\pm$ 0.00143 & 1.695 $\pm$ 0.774 & 0.05696 $\pm$ 0.02018 & 0.12202 $\pm$ 0.01045 & 60/60 \\
50 & Topology only & 0.03183 $\pm$ 0.00165 & 1.783 $\pm$ 0.872 & 0.08244 $\pm$ 0.03347 & 0.12100 $\pm$ 0.01102 & 60/60 \\
50 & Both (full) & 0.03168 $\pm$ 0.00132 & 1.455 $\pm$ 0.701 & 0.04437 $\pm$ 0.01481 & 0.12044 $\pm$ 0.00995 & 60/60 \\
\bottomrule
\end{tabularx}
\end{table*}

\section{Biological observations and evaluation tables}\label{sec:biological_validation_details}

\subsection{Observations and temporal groups}\label{sec:biological_observations_temporal_groups}

Rotifer and algae observations from chemostat C3 \cite{blasius2020} are expressed in $10^6$ cells/ml (algae), animals/ml (rotifers), and days, with the first 67 days divided into ten 6.7-day blocks (Table~\ref{tab:B1}). Monthly Least Killifish and Eastern Mosquitofish log densities at sites TP, WRGP, and WRHW \cite{gatto2024} are exponentiated and grouped into six-month blocks without interpolation across missing observations. Transitions are retained only within blocks, and blocks are assigned to training, validation, and testing to match state-space coverage while keeping model normalization restricted to the short-training starts.

\begin{table*}[!htbp]

\centering
\caption{\label{tab:B1}Temporal groups used for biological validation. Groups are numbered chronologically from zero. Longer training combines the short and additional groups, with validation and testing held fixed.}
\fontsize{9.5}{11.5}\selectfont
\setlength{\tabcolsep}{4pt}
\renewcommand{\arraystretch}{1.18}
\begin{tabularx}{\linewidth}{@{}>{\hsize=0.65\hsize\raggedright\arraybackslash}X>{\hsize=0.9\hsize\raggedright\arraybackslash}X>{\hsize=1.2\hsize\raggedright\arraybackslash}X>{\hsize=1.05\hsize\raggedright\arraybackslash}X>{\hsize=1.2\hsize\raggedright\arraybackslash}X@{}}
\toprule
\textbf{Record} & \textbf{Short training} & \textbf{Additional training} & \textbf{Validation} & \textbf{Test} \\
\midrule
C3 & 1, 7 & 2, 6, 9 & 0, 4 & 3, 5, 8 \\
TP & 0, 6 & 1, 7 & 3, 4 & 2, 5, 8, 9 \\
WRGP & 2, 4 & 3, 5 & 1, 7 & 0, 6, 8 \\
WRHW & 3, 8 & 2, 6 & 0, 5 & 1, 4, 7, 9 \\
\bottomrule
\end{tabularx}
\end{table*}

\subsection{Transition metric and summary tables}\label{sec:biological_transition_metric_longer_records}

Conditional-mean NRMSE and transition MMD$^2$ use 64 realizations integrated with four Euler--Maruyama substeps over each recorded interval $\Delta t_n$, normalized by the short-training standard deviations $s_{\mathrm{ref}}$. Transition MMD$^2$ averages three Gaussian kernels with bandwidths $0.5b, b,$ and $2b$, where $b$ is the median pairwise distance among standardized four-dimensional short-training pairs. Fish NRMSE pools standardized squared residuals across the three separately fitted sites, and fish transition MMD$^2$ averages the three site scores. Tables~\ref{tab:2} and \ref{tab:B2} report the median and interquartile range across five optimizer seeds for the short and longer training settings plotted in Figure~\ref{fig:biological_prediction_performance}.

\begin{table*}[!htbp]

\centering
{\ifshownewrevision\color{black}\fi
\caption{\label{tab:2}Test biological agreement in the short training settings. Values are median [first quartile, third quartile] over five optimizer seeds. Lower is better. For the separately fitted fish models, NRMSE pools standardized residuals and MMD$^2$ averages site scores (Section~\ref{sec:biological_records_training}).}
\fontsize{9.5}{11.5}\selectfont
\setlength{\tabcolsep}{4pt}
\renewcommand{\arraystretch}{1.18}
\begin{tabularx}{\linewidth}{@{}>{\hsize=0.8\hsize\raggedright\arraybackslash}X>{\hsize=0.95\hsize\raggedright\arraybackslash}X>{\hsize=1.05\hsize\raggedright\arraybackslash}X>{\hsize=1.05\hsize\raggedright\arraybackslash}X>{\hsize=1.15\hsize\raggedright\arraybackslash}X@{}}
\toprule
\textbf{Dataset} & \textbf{Training information} & \textbf{Model} & \textbf{Conditional mean NRMSE} & \textbf{Transition MMD$^2$} \\
\midrule
Rotifer and algae & 2 cycle blocks, 12 transitions total & TC-AdaNSDE & 0.896\newline [0.875, 0.902] & \textbf{0.0188\newline [0.0186, 0.0190]} \\
 &  & DynGMA neural SDE & 1.019\newline [1.007, 1.047] & 0.0656\newline [0.0621, 0.0765] \\
 &  & Positive neural SDE & \textbf{0.894\newline [0.868, 0.900]} & 0.0255\newline [0.0248, 0.0266] \\
 &  & Stochastic gLV & 0.968\newline [0.961, 0.973] & 0.0313\newline [0.0307, 0.0314] \\
 &  & NeuralODE & 2.656\newline [2.529, 3.082] & 0.0287\newline [0.0280, 0.0289] \\
Least Killifish and Eastern Mosquitofish & 1 year in total per site, 30 transitions total & TC-AdaNSDE & \textbf{0.966\newline [0.963, 0.976]} & 0.0265\newline [0.0256, 0.0279] \\
 &  & DynGMA neural SDE & 1.140\newline [1.081, 1.192] & \textbf{0.0239\newline [0.0238, 0.0258]} \\
 &  & Positive neural SDE & 0.989\newline [0.987, 0.994] & 0.0281\newline [0.0274, 0.0295] \\
 &  & Stochastic gLV & 1.048\newline [1.047, 1.050] & 0.0279\newline [0.0279, 0.0283] \\
 &  & NeuralODE & 1.036\newline [1.035, 1.039] & 0.0335\newline [0.0332, 0.0335] \\
\bottomrule
\end{tabularx}}
\end{table*}

\begin{table*}[!htbp]

\centering
\caption{\label{tab:B2}Predictive and distributional errors with longer biological records. Values follow the reporting conventions of Table~\ref{tab:2}. Validation and testing use the same observations as in the short settings.}
\fontsize{9.5}{11.5}\selectfont
\setlength{\tabcolsep}{4pt}
\renewcommand{\arraystretch}{1.18}
\begin{tabularx}{\linewidth}{@{}>{\hsize=0.8\hsize\raggedright\arraybackslash}X>{\hsize=0.95\hsize\raggedright\arraybackslash}X>{\hsize=1.05\hsize\raggedright\arraybackslash}X>{\hsize=1.05\hsize\raggedright\arraybackslash}X>{\hsize=1.15\hsize\raggedright\arraybackslash}X@{}}
\toprule
\textbf{Dataset} & \textbf{Training information} & \textbf{Model} & \textbf{Conditional mean NRMSE} & \textbf{Transition MMD$^2$} \\
\midrule
Rotifer and algae & 5 cycle blocks, 28 transitions total & TC-AdaNSDE & 1.250\newline [1.219, 1.253] & 0.0762\newline [0.0757, 0.0767] \\
 &  & DynGMA neural SDE & \textbf{1.059\newline [1.025, 1.127]} & 0.0768\newline [0.0511, 0.0785] \\
 &  & Positive neural SDE & 1.282\newline [1.266, 1.365] & 0.0863\newline [0.0856, 0.0889] \\
 &  & Stochastic gLV & 1.159\newline [1.152, 1.164] & 0.0655\newline [0.0651, 0.0665] \\
 &  & NeuralODE & 1.173\newline [1.163, 1.182] & \textbf{0.0564\newline [0.0527, 0.0568]} \\
Least Killifish and Eastern Mosquitofish & 2 years in total per site, 59 transitions total & TC-AdaNSDE & 0.968\newline [0.968, 0.989] & 0.0319\newline [0.0314, 0.0321] \\
 &  & DynGMA neural SDE & 1.076\newline [1.051, 1.097] & 0.0290\newline [0.0279, 0.0296] \\
 &  & Positive neural SDE & \textbf{0.963\newline [0.948, 0.982]} & \textbf{0.0288\newline [0.0280, 0.0295]} \\
 &  & Stochastic gLV & 1.035\newline [1.034, 1.037] & 0.0376\newline [0.0371, 0.0379] \\
 &  & NeuralODE & 1.081\newline [1.077, 1.082] & 0.0473\newline [0.0467, 0.0473] \\
\bottomrule
\end{tabularx}
\end{table*}

% Flush appendix floats so the references remain the final manuscript section.
\bibliography{references}
\end{document}